\documentclass[aps,twocolumn,showpacs,amsmath,amssymb]{revtex4}
\usepackage{graphicx}
\usepackage{dcolumn}
\usepackage{bm}
\usepackage{color}
\newcommand{\nn}{\nonumber}
\newcommand{\beq}{\begin{eqnarray}}
\newcommand{\eeq}{\end{eqnarray}}

\begin{document}

\title{
Vortex structures and zero energy states in the BCS-to-BEC evolution of $p$-wave resonant Fermi gases
}

\author{T. Mizushima}
\email{mizushima@mp.okayama-u.ac.jp}
\affiliation{Department of Physics, Okayama University,
Okayama 700-8530, Japan}
\author{K. Machida}
\affiliation{Department of Physics, Okayama University,
Okayama 700-8530, Japan}
\date{\today}

\begin{abstract}

Multiply quantized vortices in the BCS-to-BEC evolution of $p$-wave resonant Fermi gases are investigated theoretically. The vortex structure and the low-energy quasiparticle states are discussed, based on the self-consistent calculations of the Bogoliubov-de Gennes and gap equations. We reveal the direct relation between the macroscopic structure of vortices, such as particle densities, and the low-lying quasiparticle state. In addition, the net angular momentum for multiply quantized vortices with a vorticity $\kappa$ is found to be expressed by a simple equation, which reflects the chirality of the Cooper pairing. Hence, the observation of the particle density depletion and the measurement of the angular momentum will provide the information on the core-bound state and $p$-wave superfluidity. Moreover, the details on the zero energy Majorana state are discussed in the vicinity of the BCS-to-BEC evolution. It is demonstrated numerically that the zero energy Majorana state appears in the weak coupling BCS limit only when the vortex winding number is odd. There exist the $\kappa$ branches of the core bound states for a vortex state with vorticity $\kappa$, whereas only one of them can be the zero energy. This zero energy state vanishes at the BCS-BEC topological phase transition, because of interference between the core-bound and edge-bound states.

\end{abstract}

\pacs{05.30.Fk, 03.75.Hh, 03.75.Ss, 74.20.Rp}

\maketitle

\section{Introduction}

Spin-triplet $p$-wave superfluids, such as liquid $^3{\rm He}$, have brought rich physics due to interplay between the spin, orbital, and gauge degrees of freedom of the order parameter \cite{legett}. Various superfluid phases, so-called the ABM, BW, polar states and so on, arise from the breaking of exceptionally large symmetry group ${\rm SO}_{\bm S}(3) \!\times \! {\rm SO}_{\bm L}(3) \!\times \! {\rm U}_{\varphi}(1)$. In addition, the broken symmetry and boundary conditions give rise to the continuous configuration of the order parameter as non-trivial topological excitations, such as orbital and spin textures \cite{vollhardt,salomaa}. It has also been proposed \cite{volovik1,read} that $p$-wave superfluids can exhibit another type of the phase transition, which does not involve symmetry breaking but changes a topological charge.  The topological phase transition is driven by the enhancement of the effective pair interaction.

Recently, $p$-wave Feshbach resonances on $^{6}$Li and $^{40}$K atoms have been observed by sweeping magnetic field in experiments \cite{salomon,ketterle,jin,fuchs,inada}. The Feshbach resonance of colliding atoms into an $\ell$-wave bound state allows one to manipulate the inter-atomic interaction in $\ell$-wave channel, from the regime of weakly interacting atoms to tightly binding molecules regime. In the weakly interacting regime, the Fermi system forms Cooper pairs, where the low energy quasiparticle excitation is characterized by a $p$-wave Cooper pair potential. In contrast, the $p$-wave molecules undergo Bose-Einstein condensation (BEC), which involves the isotropic excitation gap uniquely determined by the binding energy of molecules. It has been clarified that in  contrast to the BCS-BEC crossover which appears in $s$-wave superfluids \cite{crossover}, these two regimes are not smoothly connected but give rise to the topological phase transition~\cite{volovik1,read,gurarieAP,GRA}. The experimental observation of $p$-wave Feshbach resonances is the first step towards the further realization of the $p$-wave BCS-to-BEC topological phase transition.

Here, it is natural to expect that the extremely strong coupling regime of $p$-wave superfluids contains intriguing vortex structure, because the coherence length of the order parameter $\xi$ becomes comparable to the mean interparticle distance $k^{-1}_{\rm F}$. The phase of the order parameter rotates by $2\pi\kappa$ around a quantum vortex, where $\kappa \!\in\! \mathbb{Z}$ because of its single valuedness. Hence, the quasiparticles traveling across the vortex core experience the abrupt shift of the phase, leading to the low-energy Andreev bound state \cite{tanaka} or the so-called Caroli-de Gennes-Matricon (CdGM) state \cite{cdgm}. In the regime with $\xi \!\approx k^{-1}_{\rm F}$ of $s$-wave superfluids, it has been first demonstrated in Refs.~\cite{hayashiPRL,hayashiJPSJ} that the energy levels of the CdGM state become discrete, leading to the strong depletion of the particle density around the core. The analysis on the quantum depletion has been theoretically extended to the BCS-BEC crossover regime \cite{bulgac, feder, mmachida1, mmachida2, sensarma, levin} and the depletions in the particle density were experimentally observed in rotating Fermi gases with an $s$-wave resonance \cite{mit1,mit2}. Furthermore, the structures of giant vortices with $|\kappa| \!>\! 1$ have been extensively studied in $s$-wave superfluids by a large number of authors \cite{virtanen, ktanaka, mizushima05, takahashi, hu1, hu2, suzuki}.

Turning now to $p$-wave superfluids, the CdGM states are found to be intriguing in the following two senses: (i) The low energy states in the weak coupling BCS phase are topologically distinct from them in the strong coupling BEC phase \cite{read,gurarieAP}. (ii) The lowest eigenenergy may be exactly zero in the BCS limit that the chemical potential $\mu$ is equal to the Fermi energy $E_{\rm F}$~\cite{kopnin,ivanov,volovik,yip1,fujimoto,sato}. The zero energy state (ZES) bound at the vortex core and at the edge can be described as the Jackiw-Rebbi solution in the one-dimensional Dirac equation at the domain wall \cite{jackiw,ssh,machida1,machida2,tewari,stone2}. The remarkable feature of the ZES is linked to the fact that the creation operator is identical to its own annihilation, called the Majorana fermion \cite{majorana}. In addition, since the host vortices associated with the ZES obey neither Fermi nor Bose statistics, called the non-abelian statistics \cite{ivanov,stone}, this system will offer the promising method of the fault-tolerant quantum computation \cite{review,kitaev,freedman}. It is known that they can be observed in spin-polarized $p$-wave superfluids with the time-reversal symmetry breaking, {\it e.g.}, the $k_x\pm ik_y$ chiral state. Hence, $p$-wave resonant Fermi gases can be base platform to check various theoretical issues because of the high controllability, such as two dimensionality~\cite{gurarieAP, bloch}. 

The aim of this paper is to clarify the macroscopic structures of giant vortices from the fully microscopic point of view in the BCS-BEC evolution of $p$-wave resonant Fermi gases. In the previous paper \cite{mizushima}, we revealed the roles of the ZES for visualization of $p$-wave superfluidity through the particle density depletion and net angular momentum. In this work, we expand this argument into other vortices with winding numbers $|\kappa| \!>\! 1$. Here, we use the fully microscopic theory based on the Bogoliubov-de Gennes (BdG) equation, which enables the systematic study in the BCS-to-BEC regime. It is demonstrated that the low-energy quasiparticle spectra is reflected through the quantum depletion of the particle density at the core. In addition, we calculate the net angular momentum, which is found to provide direct evidence for $p$-wave superfluidity. 
Throughout this paper, we focus on the zero temperature limit because for finite temperature regimes the pairing fluctuation effect which is not taken into account here becomes important \cite{chen}.

Furthermore, we discuss the existence of the zero energy Majorana states in the BCS-to-BEC evolution regime. It has been revealed in the BCS limit \cite{tewari} and the more generic situation \cite{gurarie} of $p$-wave superfluids that there is only a single zero energy Majorana state for odd vorticity and none for even vorticity. This is contrast to the index theorem for zero energy eigenstates of the relativistic Dirac Hamiltonian \cite{weinberg,rossi} and the quasiclassical analysis of the $p$-wave BdG equation \cite{volovik,volovik1}. Here, we reproduce the prediction and further show that the ZES in an odd-vorticity vortex disappears in the BCS-BEC topological phase transition because of the quasiparticle tunneling between the core- and edge-bound ZES's. It is also confirmed numerically that their wave functions are characterized by the modified Bessel function in the vicinity of the topological phase transition \cite{gurarie}.

In the following section, we describe the details on the self-consistent calculation based on the BdG and gap equations. In this section, we summarize the results of the vortex-free state in the BCS-to-BEC evolution. The vortex structures, such as the profiles of the order parameter and the particle density, are presented in Sec.~III. These results are discussed with the knowledge of the low-energy quasiparticle structure. Furthermore, in Sec.~IV, we demonstrate that the lowest eigenenergy becomes zero when the winding number is odd, whereas it lifts from zero in the vicinity of the BCS-BEC evolution. The final section is devoted to conclusion and discussion. The detailed derivations of the self-consistent equations and the CdGM states with arbitrary winding number are included in Appendices.

\section{Theoretical formulation}

\subsection{Self-consistent equations}

Here, we start with the mean-field Hamiltonian of spinless fermions with the mass $M$ and the Nambu spinor ${\bm \Psi}({\bm r}_1) \!\equiv\! [\psi  ({\bm r}_1), \psi^{\dag}({\bm r}_1)]^T$,
\begin{eqnarray}
\mathcal{H} = E_0 + \frac{1}{2}\int d{\bm r}_1\int d{\bm r}_2
\mbox{\boldmath $\Psi$}^{\dag}({\bm r}_1) \hat{\mathcal{K}} ({\bm r}_1,{\bm r}_2)\mbox{\boldmath $\Psi$}({\bm r}_2) ,
\label{eq:Hmf}
\end{eqnarray}
where $\psi^{\dag}$ and $\psi$ are the creation and annihilation operators of fermions. The matrix $\hat{\mathcal{K}}$ is given as
\begin{eqnarray}
\hat{\mathcal{K}} ({\bm r}_1,{\bm r}_2) = 
\left[
\begin{array}{cc}
H_0 ({\bm r})\delta({\bm r}_1-{\bm r}_2) & \Delta ({\bm r}_1,{\bm r}_2) \\
\Delta^{\ast}({\bm r}_2,{\bm r}_1) & -H^{\ast}_0 ({\bm r}_1)\delta({\bm r}_1-{\bm r}_2) 
\end{array}
\right],
\end{eqnarray}
with 
\begin{eqnarray}
H_0({\bm r}) = - \frac{\nabla^2}{2M} - \mu + i \Omega(x\partial _y - y \partial _x)
\end{eqnarray}
Throughout this paper, we set $\hbar \!=\! k_B \!=\! 1$. The rotation frequency $\Omega$ is set to be zero throughout this paper. The pair potential $\Delta ({\bm r}_1,{\bm r}_2)$ is defined as
\begin{eqnarray}
\Delta({\bm r}_1,{\bm r}_2) = - V({\bm r}_1,{\bm r}_2) \langle \psi ({\bm r}_1)\psi({\bm r}_2) \rangle.
\label{eq:delta}
\end{eqnarray}

Under $p$-wave pair potentials $\Delta _m ({\bm r})$ ($m\!=\! 0, \pm 1$), the quasiparticle eigenstate with the wave function $[u_{\nu}({\bm r}), v_{\nu}({\bm r})]^{\rm T}$ is described by the BdG equation
\begin{subequations}
\label{eq:bdgeq}
\beq
\left[
\begin{array}{cc}
H_0({\bm r}) & \Pi ({\bm r}) \\ -\Pi^{\ast}({\bm r}) & -H^{\ast}_0({\bm r})
\end{array}
\right]\left[
\begin{array}{c}
u_{\nu}({\bm r}) \\ v_{\nu}({\bm r})
\end{array}
\right] = E_{\nu}
\left[
\begin{array}{c}
u_{\nu}({\bm r}) \\ v_{\nu}({\bm r})
\end{array}
\right] ,
\eeq
\beq
\Pi ({\bm r}) = \frac{1}{k_0}\sum _{m = 0, \pm 1}
\bigg[
\Delta _{m}({\bm r})\mathcal{P}_m + \frac{1}{2}\mathcal{P}_m\Delta _m({\bm r})
\bigg],
\eeq
\end{subequations}
where we introduce the $p$-wave operators
$\mathcal{P}_{\pm 1} \!\equiv\! \mp \left( \partial _x \pm i \partial _y \right)$ and 
$\mathcal{P}_0 \!\equiv\! \partial _z$. The details on the derivation are described in Appendix A. The wave functions must satisfy the orthonormal condition
\beq
\int \left[ u^{\ast}_{\nu}({\bm r}) u_{\mu}({\bm r})
+ v^{\ast}_{\nu}({\bm r})v_{\mu}({\bm r})\right]d{\bm r}
= \delta _{\nu,\mu}.
\label{eq:normal3}
\eeq
The pair potential is self-consistently determined by
\begin{subequations}
\label{eq:gapeq}
\beq
\Delta _{\pm 1} ({\bm r}) = \frac{g_{\pm 1}}{k_0}
\bigg(\mathcal{P}^{(1)}_{\mp 1} 
- \mathcal{P}^{(2)}_{\mp 1}\bigg) \Phi({\bm r}_1,{\bm r}_2)\bigg|_{{\bm r}_2 \rightarrow {\bm r}_1} ,
\eeq
\beq
\Delta _{0} ({\bm r}) = -\frac{g_0}{k_0}
\bigg[ \bigg(\mathcal{P}^{(1)}_{0} 
- \mathcal{P}^{(2)}_{0}\bigg) \Phi({\bm r}_1,{\bm r}_2)\bigg|_{{\bm r}_2 \rightarrow {\bm r}_1} ,
\eeq
\end{subequations}
where $\Phi({\bm r}_1,{\bm r}_2)$ is defined in Eq.~(\ref{eq:phi}) with the eigenstates of Eq.~(\ref{eq:bdgeq}).

The BdG equation (\ref{eq:bdgeq}) and gap equation (\ref{eq:gapeq}) provide a qualitative formalism for the BCS-to-BEC evolution regime \cite{gurarieAP,GRA,read}, by combining with the following number equation and by using the renormalized coupling constant. First, the total particle number $N$ is fixed by adjusting the chemical potential $\mu$, where the total number is expressed as $N \!=\! \int \rho({\bm r})d{\bm r}$ with the particle density, 
\begin{eqnarray}
\rho ({\bm r}) \equiv \langle \psi^{\dag}({\bm r}) \psi({\bm r})\rangle
= \sum _{\nu} |u_{\nu}({\bm r})|^2 f(E_{\nu}).
\label{eq:rho}
\end{eqnarray}
The sum $\sum_{\nu}$ denotes the summation for all the eigenstates with the positive and negative eigenvalues. $f(E) \!=\! 1/(e^{E/T}+1)$ is the Fermi distribution function at temperature $T$. In this paper, we set temperature to be $T\!=\! 0$. 

The gap equation (\ref{eq:gapeq}) involves two divergence terms proportional to $E_{\rm c}$ and $\ln E_{\rm c}$ \cite{randeria,gurarieAP,botelho} where $E_{\rm c}$ is the cutoff energy. The dominant divergence can be removed by replacing the bare coupling constant to the renormalized one 
\begin{eqnarray}
\frac{1}{g_m} = - \frac{1}{S} \sum _{\bm k} 
\frac{\left|\Gamma ({\bm k})\right|^2}{2 \epsilon _{\bm k} - E_{\rm b}},
\label{eq:lambda}
\end{eqnarray}
with the volume of the system $S$, $\epsilon _{k} \!\equiv\! k^2/2M$, and $\Gamma ({\bm k}) \!=\! k/k_0$. Here, $E_{\rm b}$ is an eigenvalue of the Schr\"{o}dinger equation for two fermions interacting via the pairing potential $V$ \cite{randeria,botelho}. $E_{\rm b}$ is real and regarded as the two-body bound state energy in vacuum when $E_{\rm b}$ is negative, while it has an imaginary part for positive $E_{\rm b}$. However, it is known that this imaginary part is negligible in the vicinity of a $p$-wave resonance \cite{gurarieAP}. Hence, the coupling constant $g_m$ parametrized by $E_{\rm b}$ remains real and negative for all values of $E_{\rm b}$, which can remove the leading term of the ultraviolet divergence in gap equation (\ref{eq:gapeq}). Note that in the case of a two-dimensional geometry, the resulting gap equation (\ref{eq:gapeq}) with Eq.~(\ref{eq:lambda}) still contains the logarithmic divergence on $E_{\rm c}$.

\subsection{Cylindrically symmetric system}

In order to study vortex structures, it is convenient to introduce the cylindrical coordinate ${\bm r}\!=\! (r,\theta,z)$. The $\hat{k}_x \pm i \hat{k}_y$ pairs have the phase winding $\pm 1$, since $\mathcal{P}_{\pm}$ is expressed as
$\mathcal{P} _{\pm 1} \!=\! \mp e^{\pm i\theta}( \partial _r \pm \frac{i}{r} \partial _{\theta})$. We assume the cylindrical symmetry of the order parameter 
\begin{eqnarray}
\Delta _{m} ({\bm r})  = \Delta _m (r) e^{i \kappa_{m} \theta},
\end{eqnarray}
which describes the quantized vortex with a winding number $\kappa_{m} \!\in\! \mathbb{Z}$ centered at the origin $r\!=\! 0$. 

The quasiparticle wave function is then expressed in the cylindrical coordinate as
\begin{eqnarray}
\left[\begin{array}{c}
u_{\nu} ({\bm r}) \\ u_{\nu} ({\bm r})
\end{array}\right] = 
\left[ \begin{array}{c}
u_{n, \ell,q}(r)e^{i \ell _u \theta} \\  v_{n, \ell,q}(r)e^{i \ell _v \theta}
\end{array}\right]e^{iqz}.
\label{eq:qpwf}
\end{eqnarray}
Here, we impose the periodic boundary condition with the period $Z$ along the $z$-axis, i.e., $q \!=\! 2\pi n_z/Z$ with $n_z \!\in\! {\mathbb Z}$ and the axial symmetry leads to $\ell _u , \ell _v \!\in\! \mathbb{Z}$. Since we here consider a two-dimensional system, however, the axial quantum number along the $z$-axis, $q$, is set to be $q \!=\! 0$. From the BdG equation (\ref{eq:bdgeq}), we find the condition that $u_{\nu }$ couples with $v_{\nu}$ through the following angular momentum relation,
\begin{eqnarray}
\ell _u = \ell _v + \kappa _{+1} + 1 \equiv \ell, 
\label{eq:mu}
\end{eqnarray}
with the azimuthal quantum number $\ell \!\in\! {\mathbb Z}$. Also, one obtains the condition for the phase winding of $\Delta _{m}({\bm r})$ as 
\begin{eqnarray}
\kappa_{+1} = \kappa_0 -1 = \kappa _{-1} - 2 \equiv \kappa.
\label{eq:phase}
\end{eqnarray}
In addition, the following symmetry relation for eigenstates with $q\!=\! 0$ or with $\Delta _{0}\!=\! 0$ should be noted:
The positive energy state $E_{\ell}$ and $[u_{\ell}, v_{\ell}]^{\rm T}$ with $\ell$ is symmetric to the negative energy state $-E_{-{\ell}+\kappa+1}$ and $[u_{\ell}, v_{\ell}] \!=\! [v^{\ast}_{-{\ell}+\kappa+1},u^{\ast}_{-{\ell}+\kappa+1}]^{\rm T}$ with $-\ell+\kappa+1$.

\subsection{Calculated system}

In the current work, we focus on a two-dimensional system where the $Y_{1,0}(\hat{\bm k})$ orbital state of the pair potential is neglected, {\it i.e.}, $\Delta _0 ({\bm r}) \!=\! 0$. The rigid wall boundary condition is imposed on the wave functions, $u_{\nu} (r\!=\! R)\!=\! v_{\nu}(r\!=\! R) \!=\! 0$, at $R \!=\! 50 k^{-1}_F$. In this paper, we consider the strong coupling regime nearby the resonance, in which the coherence length,
\beq
\xi \equiv \frac{k_{\rm F}}{M\Delta} , \hspace{3mm} \Delta \equiv \max|\Delta _{+1}(r)|,
\eeq 
is comparable to the average of the interparticle spacing, {\it i.e.}, $\xi \!\sim\! k^{-1}_F$. Hence, the radius $R$ which we set here is enough to recover the pairing field from the vortex center, $R\!\gg\!\xi$. The resulting system consists of the $N \!=\! 600$ fermions in a single hyperfine spin state. All the quantities are scaled by using the length unit $k^{-1}_{\rm F}$ and the energy unit $E_{\rm F}$, where $E_{\rm F} \!=\! k^2_{\rm F}/2M$ is the Fermi energy. The energy cutoff is set to be $E_{\rm c} \!=\! 60E_{\rm F}$. We also assume $k_0 \!=\! k_{\rm F}$.

In the vortex-free state of two-dimensional spinless Fermi systems, the $\hat{k}_x \pm i \hat{k}_y$ pairing states are energetically degenerate, called the orbital ferromagnetic or chiral state. The non-zero vorticity, however, breaks the degeneracy and vortex states with a positive winding number becomes distinguishable from the negative winding states. Hereafter, without the loss of generality, we consider the system that $\Delta _{+1}$ is dominant.

\subsection{Vortex-free state}

\begin{figure}[t!]
\includegraphics[width=80mm]{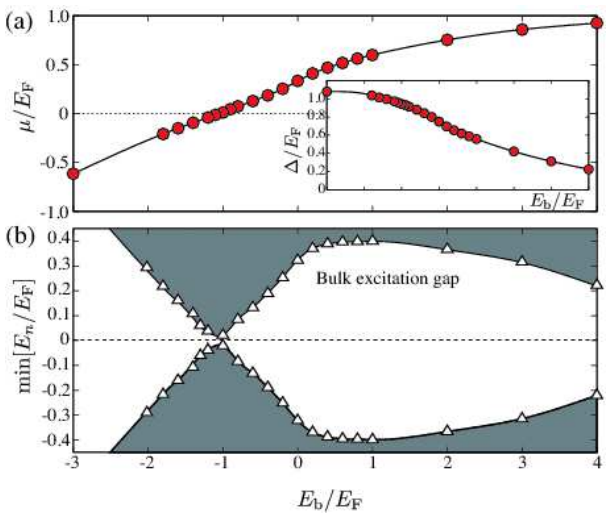}
\caption{(Color online) (a) Chemical potential $\mu$ as a function of $E_{\rm b}$ in the vortex-free state with $\kappa \!=\! 0$. The maximum value of the pair potential $\Delta/E_{\rm F} \!=\! \max|\Delta _{+ 1}(r)|/E_{\rm F}$ is plotted as functions of $E_{\rm b} \!\in\! [-3,4]$ in the inset of (a). (b) Bulk excitation gap in the vortex-free state as a function of $E_{\rm b}$. 
}
\label{fig:cp}
\end{figure}

First of all, we present in Fig.~\ref{fig:cp} the numerical results of the self-consistent equations (\ref{eq:bdgeq}) and (\ref{eq:gapeq}) with Eq.~(\ref{eq:lambda}) in the vortex-free state ($\kappa \!=\! 0$). The chemical potential $\mu$ and the maximum value of the pair potential $\Delta \!\equiv\! \max{|\Delta _{+1}(r)|}$ are plotted as a function of $E_{\rm b}$ in Fig.~\ref{fig:cp}(a). Feshbach resonance allows one to manipulate the effective coupling constant $g_{\pm 1}$, where the Cooper pairs realized within $\mu \!>\! 0$, say the BCS phase, turn to molecular bosons in the BEC phase when $\mu \!<\! 0$ \cite{gurarieAP,GRA,read}. The coupling constant $g_m$ is parameterized with $E_{\rm b}$ through Eq.~(\ref{eq:lambda}). In $E_{\rm b}\!>\! 0$, $\mu$ approaches the Fermi energy $E_{\rm F}$ with increasing $E_{\rm b}$. It is seen from the inset of Fig.~\ref{fig:cp}(a) that throughout the whole range of $E_{\rm b}$, the amplitude of the pair potential $\Delta$ yields no singular behavior. 

Figure~\ref{fig:cp}(b) shows the low-energy quasiparticle spectrum in the bulk. In the BCS limit $E_{\rm b} \!\gg\! E_{\rm F}$, the lowest excitation gap in the bulk is characterized by the dissociation energy of the Cooper pair $\min|E|\!=\! \Delta $. In the vicinity of $|\mu| \!\ll\! E_F$, however, the energy gap turns to $\min|E| \!=\! |\mu|$, which indicates that the excitation becomes gapless. In our calculation, this occurs at $E_{\rm b}/E_{\rm F} \!\approx\! -1.0$, as seen in Fig.~\ref{fig:cp}(b). Read and Green \cite{read} predicted that the point with $\mu \!=\! 0$ separates two topologically distinguishable phases, such as the BCS ($E_{\rm b}/E_{\rm F}\!>\! -1.0$) and BEC ($E_{\rm b}/E_{\rm F}\!<\! -1.0$) phases. This is contrast to the case of the $s$-wave pairing; The excitation spectrum in the $\mu \!>\! 0$ regime has a gapful with $|\Delta|$, which continuously turns to $\min|E| \! =\! \sqrt{|\mu|^2 + |\Delta|^2}$ as $\mu$ becomes negative across $\mu \!=\! 0$, {\it i.e.}, the BCS-to-BEC crossover \cite{crossover}. 

\section{Vortex structure}

\subsection{Pair potential}

In two-dimensional systems, the order parameter reduces to two components, $\Delta _{\pm 1}$, because of $q \!=\! 0$. Here, we consider vortex states that the majority component $\Delta _{+1}$ has a winding number $|\kappa | \!\le\! 3$. As seen in Eq.~(\ref{eq:phase}), the axial symmetry of the system requires the winding number to satisfy the condition $\kappa _{+1} \!=\! \kappa _{-1} -2 \!\equiv\! \kappa$. Hence, we take account of 6 combinations of winding numbers, $\langle \kappa _{+1}, \kappa _{-1} \rangle \!=\! \langle -3, -1 \rangle$, $\langle -2, 0 \rangle$, $\langle -1, 1 \rangle$, $\langle 1, 3 \rangle$, $\langle 2, 4 \rangle$, and $\langle 3, 5 \rangle$. Here, we shall discuss all the vortex states, except for $\langle \kappa _{+1}, \kappa _{-1}\rangle \!=\! \langle -1, 1 \rangle$, because its winding configuration was studied in our previous work \cite{mizushima}. 

\begin{figure}[b!]
\includegraphics[width=80mm]{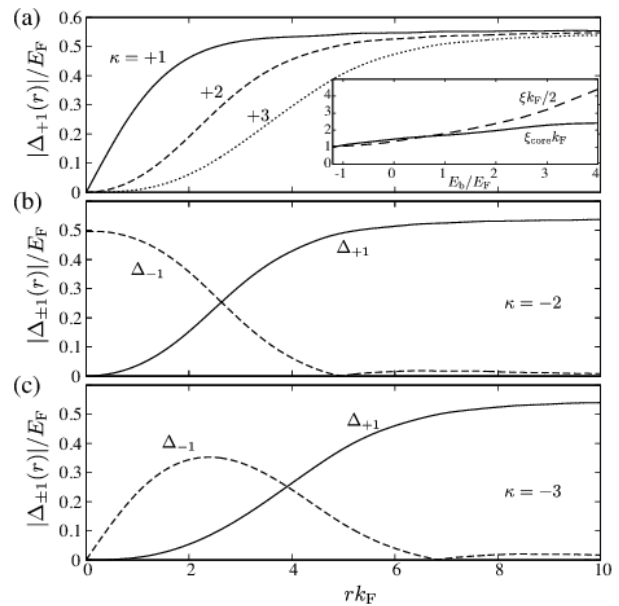}
\caption{(a) Profiles of the pair potential $\Delta _{+1} (r)$ in $\kappa  \!=\! +1$ (solid line), $+2$ (dashed line), $+3$ (dotted line) at $E_{\rm b} / E_{\rm F} \!=\! 1.0$. (b,c) Profiles of pair potentials $\Delta _{+1} (r)$ (solid line) and $\Delta _{-1} (r)$ (dashed line): (b) $\kappa  \!=\! -2$ and (c) $\kappa \!=\! -3$ at $E_{\rm b}/E_{\rm F} \!=\! 1.0$. The inset in (a) shows the vortex radius $\xi_{\rm core}$ normalized by $k^{-1}_{\rm F}$ as a function of $E_{\rm b}$. The solid and dashed lines denote $\xi _{\rm core}$ and $\xi/2$, respectively.
}
\label{fig:delta}
\end{figure}

Figure~\ref{fig:delta} shows the profile of the pair potentials $\Delta _{\pm 1} (r)$ near the vortex center $r\!=\! 0$ in the $\kappa  \!=\!  -2, -3, 1, 2, 3$ vortex state. Here, it is convenient to categorize the vortex states in terms of the sign of $\kappa$. The ``positive'' (``negative'') vortex state is defined as the state that the vorticity $\kappa $ is parallel (anti-parallel) to the chirality of the orbital motion $k_x+ik_y$ of pairs. In general, a quantum vortex is defined as the phase singularity at which the amplitude of the order parameter vanishes as $\Delta (r \!=\! 0)=0$. The order parameter recovers to the constant value within the coherence length $\xi$. The slope of $\Delta (r)$ near $r\!=\! 0$ is characterized by the winding number, {\it i.e.}, $\lim _{r \!\rightarrow\! 0} \Delta _m (r) \!\propto\! r^{|\kappa _m|}$. It is seen in Fig.~\ref{fig:delta}(a) that in the case of the positive vortex state, the winding number $\kappa _m$ determines the size of the vortex core, which becomes larger as $\kappa _m$ increases. The vortex core of $\Delta _{+1}$ in the case of the positive vortex state is always smaller than that in the induced component $\Delta _{-1}$, because of $\kappa _{+1} \!<\! \kappa _{-1}$. Hence, the giant cores of $\Delta _{+1}$ leave empty and the resulting pair potential is effectively describable with the single component $\Delta _{+1} (r)$, where $\Delta _{-1}(r)$ with the larger vortex core is suppressed. In addition, the core size of the $\kappa \!=\! +1$ vortex can be quantified as $\xi^{-1}_{\rm core} \!=\!\lim _{r \!\rightarrow \! 0}|\bar{\Delta}_{+1}(r)|/r$ with $\bar{\Delta}_{+1}(r) \!=\!\Delta _{+1}(r)/\Delta$ \cite{hayashiPRL}. As seen in the inset of Fig.~\ref{fig:delta}(a), the strong coupling effect gradually shrinks the core radius $\xi _{\rm core}$, where the vortex core size is of the order of the atomic scale $\approx k^{-1}_{\rm F}$. 

In contrast, since $|\kappa _{+1}| \!>\! |\kappa _{-1}|$ in the case of the negative vortex states with $\kappa _{+1} \!<\! -1$, the induced component $\Delta _{-1}$ always has the vortex core smaller than that of $\Delta _{+1}$. It is seen in Figs.~\ref{fig:delta}(b) and \ref{fig:delta}(c) that the $\Delta _{-1}$ component is induced inside of the vortex core of majority $\Delta _{+1}(r)$. In particular, the vortex core of $\Delta _{+1}$ with $\kappa \!=\! -2$ is filled in by the induced component $\Delta _{-1}$ which is the vortex-free state with $\kappa _{-1} \!=\! 0$. Hence, the quasiparticle structure at the core exhibits gapful. In the same sense, since the vortex structure of $\kappa \!=\! -3$ in the vicinity of the origin is dominated by the induced component, the quasiparticle structure and particle density at the core are found to be almost same as those in the single-vortex state of $\Delta _{-1}$ with $\kappa _{-1} \!=\! -1$. 

Note that these vortex structures and the local density of states are also discussed in Ref.~\cite{sauls} within the quasiclassical approximation, corresponding to the BCS limit with $\mu \!=\! E_{\rm F}$. In the BCS regime, {\it e.g.}, $E_{\rm b} /E_{\rm F} \!=\! 1.0$, the pair potentials of all the vortex configurations which we obtain here is found to be consistent with those in Ref.~\cite{sauls}. 

\subsection{Low-lying spectra}

\begin{figure*}[t!]
\includegraphics[width=170mm]{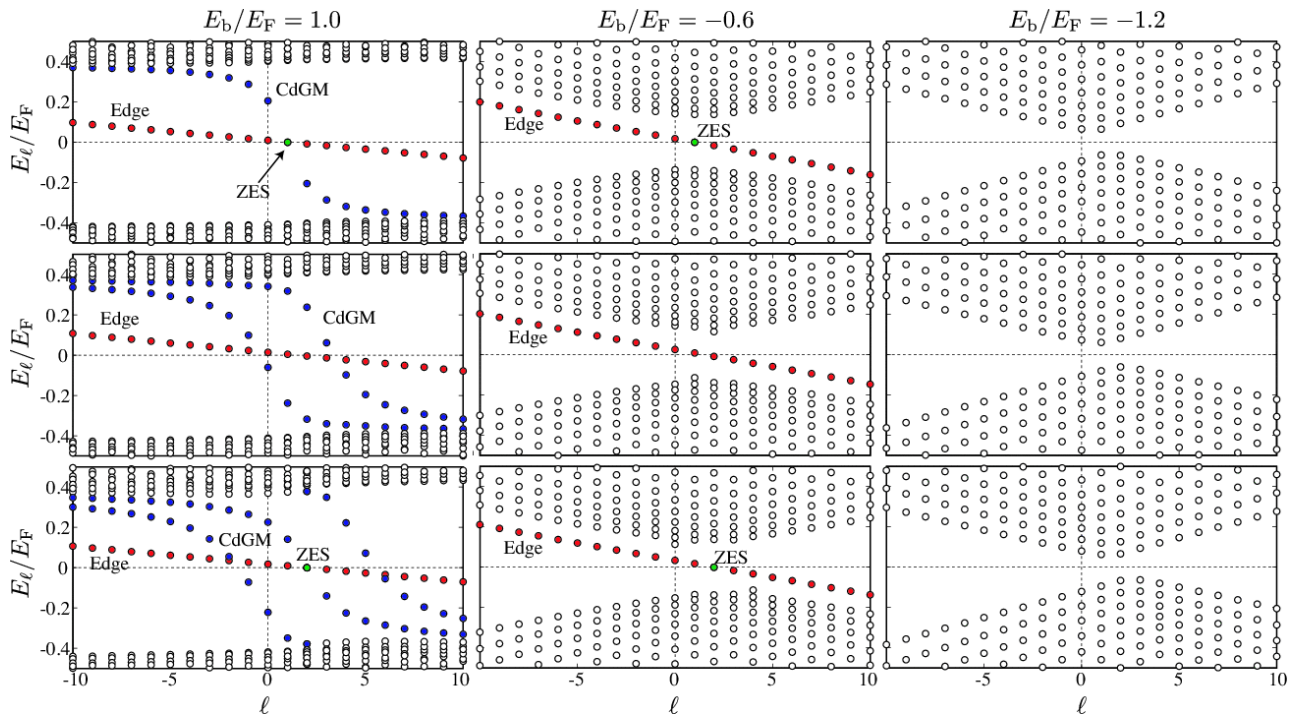}
\caption{(Color online) Quasiparticle excitation spectra at $E_b/E_F \!=\! 1.0$ (left column), $-0.6$ (center column), and $-1.2$ (right column) in the vortex states with $\kappa  \!=\! 1$ (top row), $\kappa  \!=\! 2$ (middle row), and $\kappa \!=\! 3$ (bottom row). ``CdGM'', ``Edge'', and ``ZES'' denote the branches of the CdGM, edge, and zero energy state, respectively.
}
\label{fig:spectra}
\end{figure*}

We draw attention to the quasiparticle excitation spectrum under the pair potential displayed in Fig.~\ref{fig:delta}. In particular, we focus on the vortex states with positive winding numbers $\kappa  \!=\! +1, +2, +3$. In Fig.~\ref{fig:spectra}, we plot the low-lying energy spectra of their vortex states at three regimes: The weak coupling BCS regime ($E_{\rm b} /E_{\rm F} \!=\! 1.0$), the vicinity of the BCS-to-BEC transition point ($E_{\rm b} /E_{\rm F} \!=\!-0.6$), and the BEC regime ($E_{\rm b} /E_{\rm F} \!=\!-1.2$). It is found that two branches are embedded inside the bulk energy gap $|E|\!<\!0.4E_{\rm F}$ in the BCS regime of $\kappa \!=\! 1$. The low-lying branch labeled as the ``Edge'' in Fig.~\ref{fig:spectra} consists of the eigenstates whose wave functions are localized in the edge of the system. This edge mode branch exists in the other vortex state, {\it e.g.}, $\kappa \!=\! 2$ and $3$, as seen in Fig.~\ref{fig:spectra}. The eigenenergy of this edge mode in the $k_x + ik_y$ pairing state with a vortex winding number $\kappa $ obeys the linear dispersion on the azimuthal quantum number $\ell$ \cite{stone2}, 
\beq 
E_{\ell} \!=\! - \left(\ell - \frac{\kappa +1}{2}\right)\epsilon,
\label{eq:edge}
\eeq 
where the energy spacing is found to be $\epsilon \!\approx\! \Delta /k_{\rm F}R \!=\! 0.008 E_{\rm F}$ at $E_{\rm b}/E_{\rm F} \!=\! 1.0$. Note that the eigenenergy of the edge mode can become zero at $\ell \!=\! \frac{\kappa +1}{2}$, when $\kappa $ is odd. 

In the weak coupling regime ($E_{\rm b}/E_{\rm F} \!=\! 1.0$), the eigenstate belonging to the other branch inside the bulk excitation gap $E \!=\! \pm \Delta  \!\approx\! 0.4E_{\rm F}$ is identified as the core bound state, {\it i.e.}, the CdGM state labeled as the ``CdGM'' in Fig.~\ref{fig:spectra}. The eigenenergy of the CdGM branch in the $k_x+ik_y$ pairing system with arbitrary winding number $\kappa \!\neq\! 0$ can be analytically expressed with $n \!\in\!\mathbb{Z}$ and $q\!=\! 0$ as 
\begin{eqnarray}
E_{\ell,n} = - \left( \ell - \frac{\kappa +1}{2}\right) \omega _0
+ \left( n - \frac{\kappa +1}{2} \right) \omega _1. 
\label{eq:cdgm}
\end{eqnarray}
This is obtained from the BdG equation (\ref{eq:bdgeq}) by extending the procedure by Caroli {\it et al.} \cite{cdgm} to the chiral $p$-wave system, which is valid for the eigenstate within $|\ell|\!\ll\! k_F\xi$. The expressions of two coefficients $\omega _{0,1}$ are described in Eq.~(\ref{eq:analyticE}). As seen in Eqs.~(\ref{eq:cdgm}) and (\ref{eq:analyticE}), even though the minority component $\Delta _{-1}(r)$ contributes to the coefficients $\omega _{0,1}$, the eigenenergy in Eq.~(\ref{eq:cdgm}) maintains the zero energy state with $\ell \!=\! \frac{\kappa + 1}{2}$ when $\kappa$ is odd. Since these are approximated as $\omega _0 \!\approx\! \frac{1}{2}\frac{\Delta^2}{E_F}$ and $\omega _1 \!\approx\! \frac{\pi}{2}\Delta$ in the BCS limit ($\Delta \!\ll\! E_F$), one reads $\omega _0 \!\ll\! \omega _1$. The numerical results in Fig.~\ref{fig:spectra} reproduce the linear behavior on $\ell$. The wave function is exponentially localized inside the core:
\beq
\left[
\begin{array}{c}
u_{\ell} ({\bm r}) \\
v_{\ell} ({\bm r})
\end{array}
\right] = \mathcal{N} e^{i\ell\theta}
\left[
\begin{array}{c}
f_{\ell,n}(k_{+}r) \\
f_{\ell-\kappa-1,n}(k_{-}r)e^{-i(\kappa+1)\theta}
\end{array}
\right],
\label{eq:cdgmwf}
\eeq 
where $\mathcal{N}$ is the normalization constant and 
\beq
f_{\ell,n}(k_{\pm}r) \equiv J_{\ell}(k_{\pm}r) e^{-\frac{M}{k_{\rm F}}\int^{r}_0[\Delta _{+1}(r')-\Delta _{-1}(r')]dr'}.
\label{eq:cdgmwf2}
\eeq
Here, $k_{\pm} \!\equiv\! k_{\mu}\pm ME_{\ell,n}/k_{\mu}$. Note that $k_{\pm} \!\approx\! k_{\mu}$ for the low-lying eigenstates in the BCS regime. The further details are described in Appendix B. As the strong coupling regime is approached, however, the CdGM wave function spreads over the whole region of the system. This will be discussed in Sec.~IV. 

It is found that the number of the CdGM branches uniquely depends on the winding number $\kappa$ in the weak coupling BCS regime, {\it e.g.}, three CdGM branches appear in the $\kappa \!=\! +3$ vortex state. In addition, the lowest eigenenergy of both the CdGM and edge states in Eqs.~(\ref{eq:edge}) and (\ref{eq:cdgm}) can be zero only when $\kappa $ is odd, which is numerically demonstrated in Fig.~\ref{fig:spectra}. This reproduces the prediction derived in Refs.~\cite{tewari,gurarie} that there is only single zero energy state for odd vorticity and none for even $\kappa$. It is obvious that the CdGM wave function with $\ell$ exhibits $u_{\ell}(r) \!\propto\! J_{\ell}(k_{\mu}r)\!\approx\! r^{|\ell|}$ and $v_{\ell}(r) \!\approx\! r^{|\ell - \kappa -1|}$ at $r\!\rightarrow\! 0$. Hence, the asymptotic wave functions of the zero energy states with $\ell \!=\! (\kappa+1)/2 \!\in\! \mathbb{Z}$ exhibits $u_{\ell} (r) \!=\! v_{\ell}(r) \!\approx\! r^{|\kappa +1|/2}$. The further details on the zero energy states will be described in Sec.~IV. 

In the vicinity of the BCS-BEC phase transition $|\mu| \!\ll\! E_{\rm F}$, {\it e.g.}, $E_{\rm b}/E_{\rm F} \!=\! -0.6$, the bulk energy gap is characterized as $\min|E_{\rm bulk}| \!=\! |\mu|$, as seen in Fig.~\ref{fig:cp}. 
In this regime, the amplitude of the pair potential is comparable to the Fermi energy, $\Delta  \!\approx\! E_{\rm F}$. This means that the energy level distance of the CdGM states becomes comparable to the Fermi energy, {\it i.e.}, $\omega _0 \!\approx\! E_{\rm F}$ in Eq.~(\ref{eq:cdgm}). In contrast, the bulk excitation gap, {\it i.e.}, the lowest energy of the continuum states, becomes proportional to the amplitude of the chemical potential, $\min|E_{\rm bulk}| \!=\! |\mu|$. Since $|\mu| \!<\!E_{\rm F}$ in this regime, the CdGM states are merged to the continuum state with the bulk excitation gap $\min|E_{\rm bulk}| \!=\! |\mu| \!<\! \omega _0 \!\approx\! \Delta^2/E_{\rm F}$.
Hence, as seen in $E_{\rm b}/E_{\rm F} \!=\!-0.6$ of Fig.~\ref{fig:spectra}, the low-lying spectra in this regime consist of the branch of the edge state with or without the zero energy state and bulk excitation gap. Note that the bulk excitation gap $|\mu|$ is still larger than the level distance of the edge mode $\epsilon$ in Eq.~(\ref{eq:edge}) which is determined by the inverse of the radius of the system. 

Beyond the topological phase transition $E_{\rm b}/E_{\rm F} \!=\! -1.0$, the low-lying spectra becomes trivial, where the excitation gap is uniquely characterized by $\min|E_n| \!=\! |\mu|$. It is seen from Fig.~\ref{fig:spectra} that the branch of the edge state and the zero energy state disappear in the BEC phase.

\begin{figure}[b!]
\includegraphics[width=75mm]{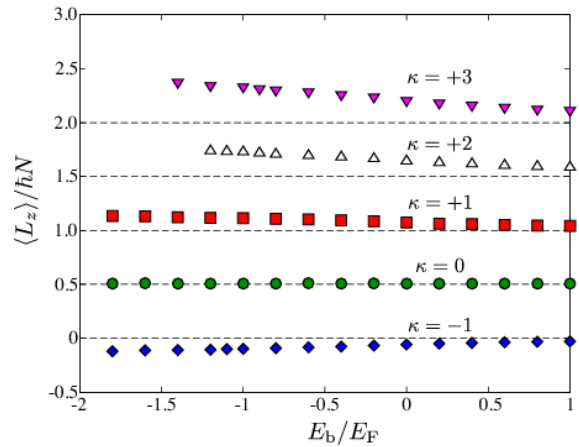}
\caption{(Color online) Total angular momentum $\langle L_z \rangle$ as a function of $E_{\rm b}$ in various vortex states with $\kappa  \!=\! -1, 0, +1, +2, +3$. The dashed lines denote Eq.~(\ref{eq:lz}).
}
\label{fig:lz}
\end{figure} 

However, it should be emphasized that even in the BEC phase, the spectra is asymmetric with respect to $\ell$. This arises from the fact that the BdG equation (\ref{eq:bdgeq}) still requires the one-to-one mapping between $[u_{\ell}, v_{\ell}]^{\rm T}$ with $E_{\ell}$ and $[u^{\ast}_{-\ell+\kappa +1}, v^{\ast}_{-\ell+\kappa +1}]^{\rm T}$ with $-E_{-\ell+\kappa +1}$. This asymmetry on $\ell$ gives rise to the non-trivial angular momentum. The total angular momentum per particle is estimated by the fully self-consistent calculation as 
\beq
\frac{\langle L_z \rangle}{\hbar N} \!\approx\! \frac{\kappa +1}{2},
\label{eq:lz}
\eeq
in the whole range of $E_{\rm b}$ in the $k_x + ik_y$ pairing state, as seen in Fig.~\ref{fig:lz}. The deviation of the self-consistently obtained $\langle L_z \rangle$ from Eq.~(\ref{eq:lz}) arises from the vortex winding and chirality of the induced component $\Delta _{-1}(r)$. In the weak coupling BCS limit, the total angular momentum per a particle $\langle L_z \rangle/ N$ consists of the angular momentum due to the vorticity $\hbar\kappa/2$ in addition to the chirality of the $k_{x}+ik_y$ channel $+\hbar/2$, where the latter originates from the linear dispersion of the edge state in the BCS regime \cite{stone2,stone08}. Figure \ref{fig:lz} demonstrates that this argument can be expanded to giant vortex states in the vicinity of the BCS-BEC evolution. Note that since $\langle L_z \rangle/ N \!\approx\! \hbar\kappa /2$ in $s$-wave superfluids, the measurement of the total angular momentum through the shift of quadrupole frequencies provides direct evidence for $p$-wave superfluidity in experiments \cite{mizushima}.

\subsection{Particle density depletion and local density of states at vortex cores}

The direct observation of the low-lying quasiparticle spectra is experimentally challenging problem. Nevertheless, it has been discussed in $s$-wave superfluids with arbitrary winding number \cite{hayashiJPSJ,bulgac,levin,sensarma,mizushima05,takahashi,suzuki,hu1,hu2,mmachida1,mmachida2} and $p$-wave superfluids with $|\kappa | \!\le\! 1$ \cite{matsumoto2,mizushima} that the local particle density $\rho (r)$ around the vortex core reflects the low-lying CdGM  spectrum. 

We here extend this analysis to giant vortex states in $p$-wave BCS-BEC evolution. Among the various possible vortices, the particle density depletion in the $\kappa  \!=\! \pm 1$ vortex state was discussed in the weak coupling regime \cite{matsumoto2}. In addition, the $\kappa \!=\! -1$ vortex in the vicinity of the BCS-to-BEC evolution was studied in Ref.~\cite{mizushima}. It is found that the core of the $\kappa \!=\!-1$ vortex in the weak coupling BCS regime $E_{\rm b} \!\gg\! E_{\rm F}$ is invisible through the density profile, which reflects the fact that the vortex core is filled in by the zero energy CdGM state with $\ell \!=\! 0$, {\it i.e.}, the Majorana state. As the BCS-BEC topological phase transition point is approached, however, since the zero energy state lifts from zero to finite energy, the vortex gradually becomes visible. As a result, the quantum depletion of the $\kappa  \!=\! -1$ vortex turns out to be closely associated with the instability of the zero energy state with $\ell \!=\! 0$ across the BCS-BEC topological phase transition point. 

\begin{figure}[b!]
\includegraphics[width=80mm]{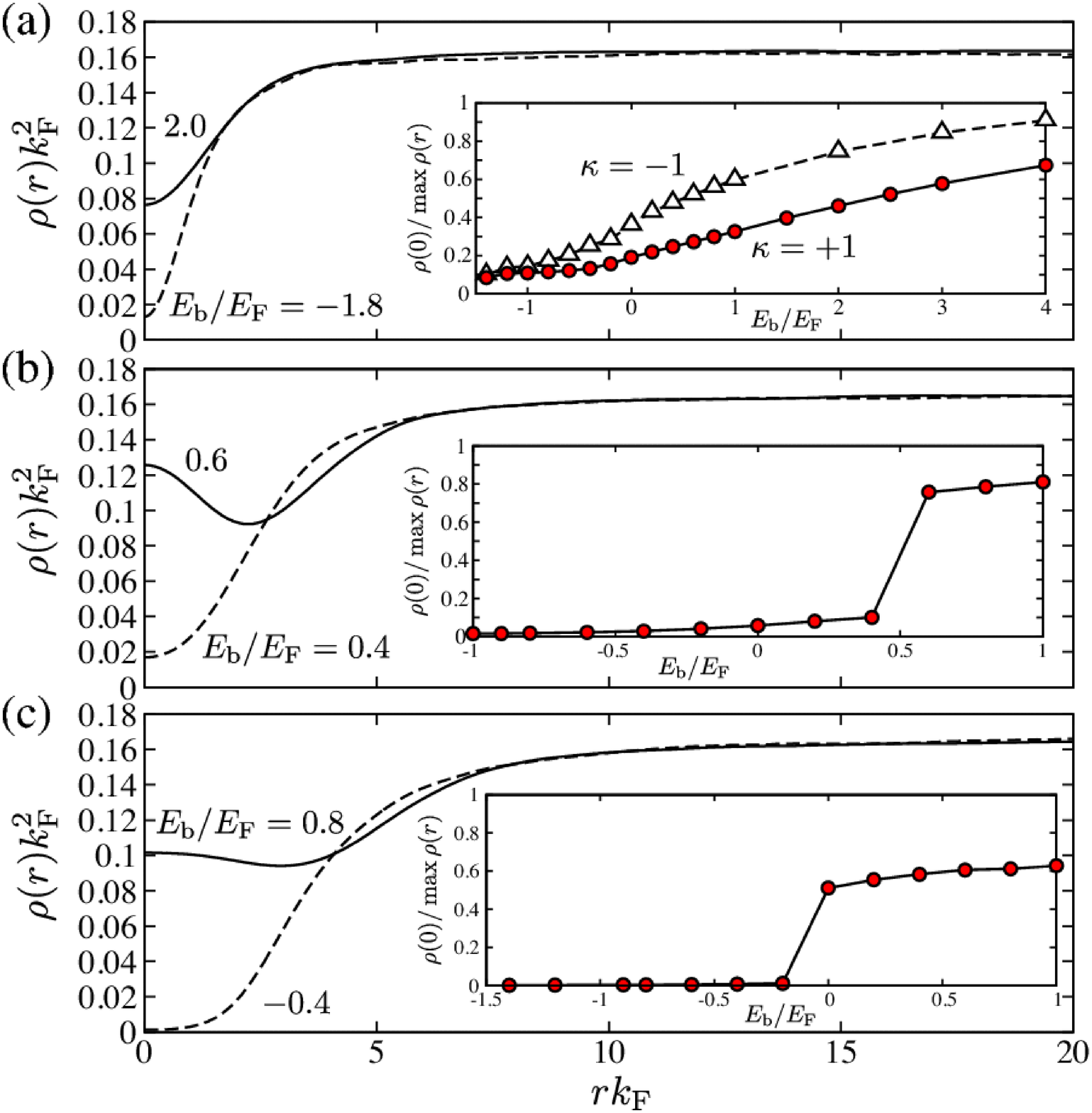}
\caption{(Color online) Spatial profiles of the particle density $\rho (r)$ in $\kappa  \!=\! +1$ vortex state (a), $\kappa \!=\! +2$ (b), and $\kappa \!=\! +3$ (c). The insets in (a)-(c) show the density depletion $\rho(r\!=\! 0)/\max\rho(r)$ at the vortex core as a function of $E_{\rm b}$. For comparison, we plot $\rho(r\!=\! 0)$ in the $\kappa  \!=\! -1$ vortex state with open triangles in the inset of (a).
}
\label{fig:dns}
\end{figure}

Let us now attempt to expand this argument into the vortex states with the other $\kappa$. Figure~\ref{fig:dns} shows the particle density profile around the giant vortex cores with  $\kappa  \!=\! +1$, $+2$, and $+3$. In the weak coupling limit $E_{\rm b} \!\gg\! E_{\rm F}$ of all the vortex states, the vortex core region is filled in by particles, while $\rho(r \!=\! 0)$ becomes zero in the BEC limit $E_{\rm b}/E_{\rm F} \!\ll\! -1$. In the $\kappa  \!=\! +1$, the particle density at the core $\rho(r \!=\! 0)/\max\rho(r)$ is plotted in the inset of Fig.~\ref{fig:dns}(a), which shows smooth suppression toward zero as $E_{\rm b}$ decreases. Here, the core of $\kappa  \!=\! +1$ vortex is quantitatively more visible than that of the $\kappa  \!=\! -1$ vortex  in the whole range of $E_{\rm b}$. In the giant vortex states with $\kappa  \!=\! +2$ and $+3$, however, it is seen from the inset of Figs.~\ref{fig:dns}(b) and \ref{fig:dns}(c) that discontinuity appears in the depletion at certain $E_{\rm b}$, {\it e.g.}, $E_{\rm b} /E_{\rm F} \!\approx\! 0.5$ for $\kappa  \!=\! +2$ and  $E_{\rm b} /E_{\rm F} \!\approx\! -0.1$ for $\kappa  \!=\! +3$.

These behaviors on the particle density, such as the depletion and its discontinuity, are closely linked to the CdGM state with $\ell \!=\! 0$. In the BCS regime, we have seen in the previous section that the low-energy excitation spectra consist of the CdGM and edge states in addition to the bulk excitation gap. Hence, it is natural to decompose the particle density around the core into two contributions: $\rho(r) \!=\! \rho _{\rm CdGM}(r) + \rho _{\rm bulk}(r)$. Here $\rho _{\rm bulk}(r) \!\equiv\! \sum _{|E|\!>\! \Delta _0} |u_{\ell}(r)|^2f(E)$ may be assumed to be uniform in this regime, while one finds 
\beq
\rho _{\rm CdGM} (r\!\rightarrow\! 0) \!=\! \sum _{\ell \ge \frac{\kappa+1}{2}} |u_{\ell}(r\!\rightarrow\! 0)|^2
\approx \sum _{\ell\ge\frac{\kappa +1}{2}} r^{2|\ell|},
\label{eq:rcdgm}
\eeq 
with Eqs.~(\ref{eq:cdgm}) and (\ref{eq:cdgmwf}). One finds that $\rho _{\rm CdGM} (r\!\rightarrow\! 0) \!\approx\! \sum _{\ell \!>\! 0}r^{2|\ell|}\!\approx\! r^0$ for the $\kappa  \!=\! -1$ vortex and $\rho _{\rm CdGM} ((r\!\rightarrow\! 0) \!\approx\! \sum _{\ell \!>\! 1}r^{2|\ell|} \!\approx\! r^2$ for the $\kappa  \!=\! +1$. Hence, this implies that the quantum depletion displayed in the inset of Fig.~\ref{fig:dns}(a) reflects the discrepancy of $\rho _{\rm CdGM}(r \!=\! 0)$ between $\kappa  \!=\! \pm 1$. 

Following the same argument, the giant vortices with the larger winding number are also understandable. In the weak coupling limit $\Delta^2_0/E_{\rm F}\!\ll\! 1$, it is found from Eq.~(\ref{eq:cdgm}) that two (three) CdGM states with $\ell\!=\! 0$ can exist inside the bulk gap in the $\kappa  \!=\! +2$ ($+3$) vortex state. For instance, $E_{\ell \!=\! 0} \!\approx\! \pm \frac{\Delta }{2} + \mathcal{O}(\Delta^2/E_{\rm F})$ is obtained from Eq.~(\ref{eq:cdgm}) in the case of $\kappa  \!=\! +2$. Only the negative energy state contributes to $\rho _{\rm CdGM}(r\!=\! 0)$ in Eq.~(\ref{eq:rcdgm}). However, it is predicted that the correction with $\mathcal{O}(\Delta^2/E_{\rm F})$ shifts the energy $E_{\ell \!=\! 0}$ upward as the strong coupling regime is approached. As a result, the shift of $E_{\ell \!=\! 0}$ from the negative to positive energy leads to the abrupt depletion of $\rho (r \!\approx\! 0)$. This argument is also applicable to the case of $\kappa \!=\! 3$, where the three lowest eigenenergies of the $\ell \!=\! 0$ CdGM state are found to be $E_{\ell \!=\! 0} \!\approx\! \mathcal{O}(\Delta^2/E_{\rm F})$ and $E_{\ell \!=\! 0} \!\approx\! \pm \Delta  + \mathcal{O}(\Delta^2/E_{\rm F})$. Only the distinction between $\kappa \!=\! +2$ and $+3$ arises from the quantitative difference of $E_{\ell \!=\! 0}$, resulting in the fact that the critical value of $E_{\rm b}$, at which $\rho(r\!=\! 0)$ exhibits jump seen in the inset of Figs.~\ref{fig:dns}(b) and \ref{fig:dns}(c), depends on $\kappa$.

\begin{figure}[b!]
\includegraphics[width=85mm]{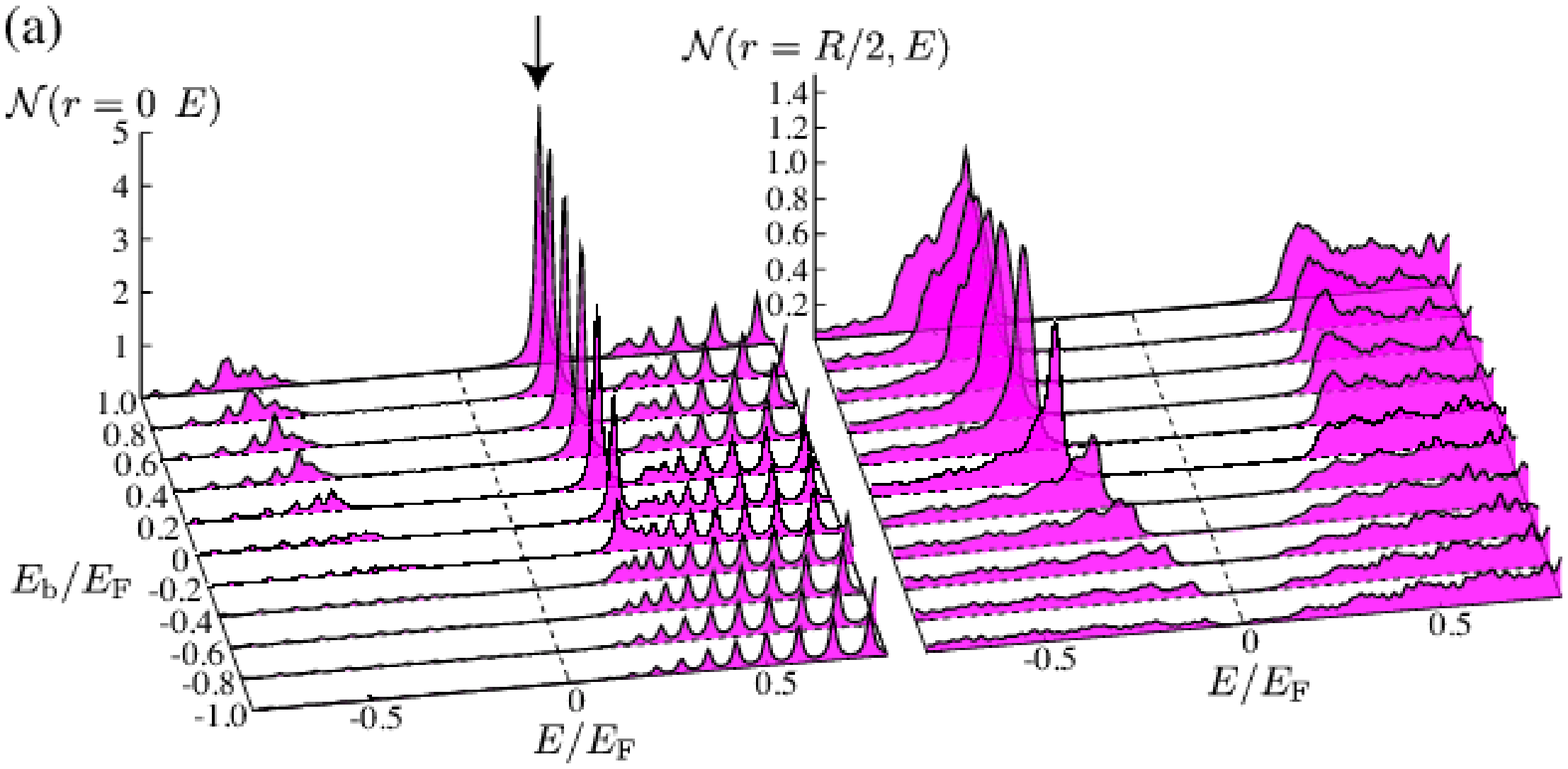}
\includegraphics[width=85mm]{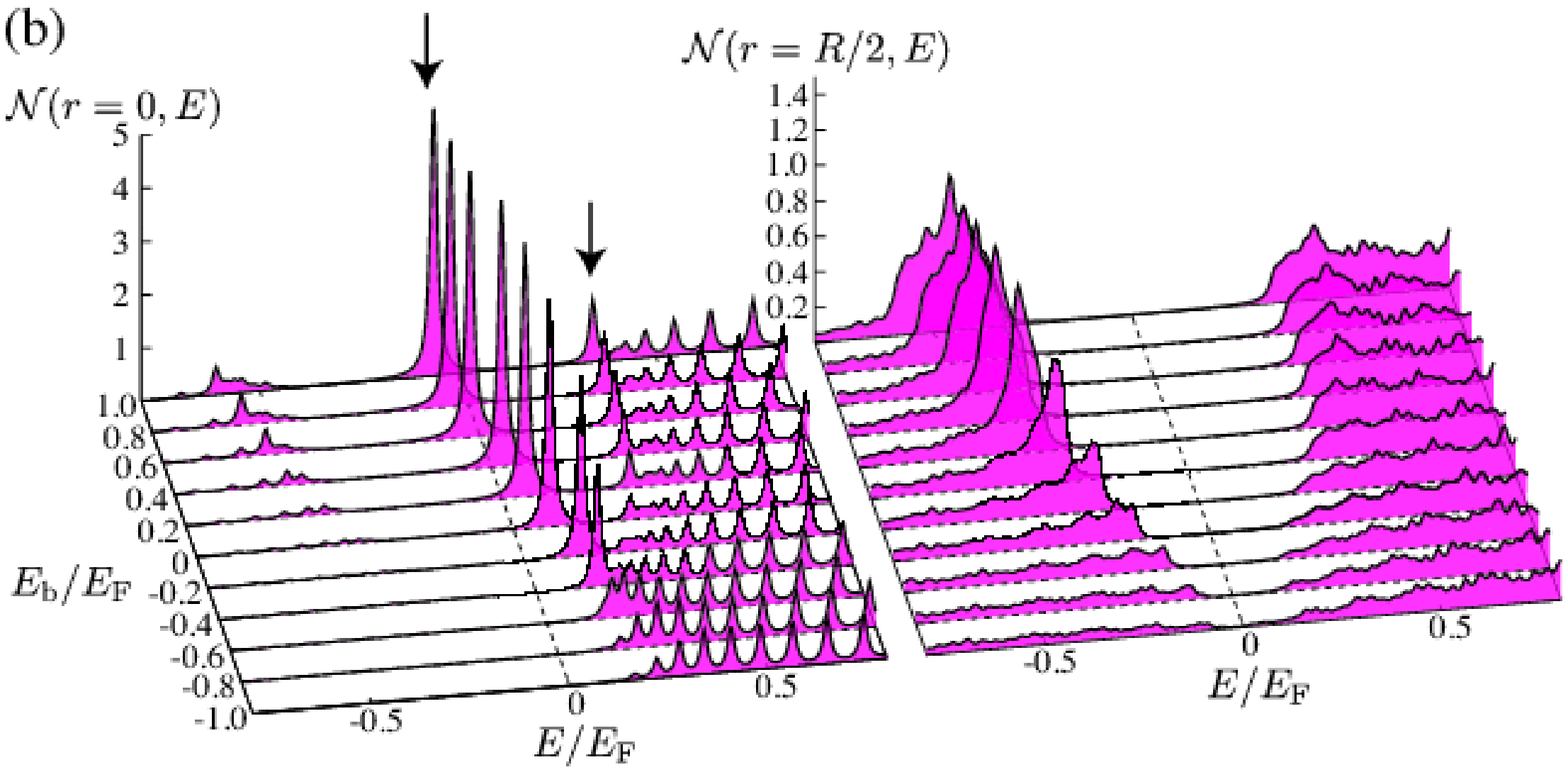}
\includegraphics[width=85mm]{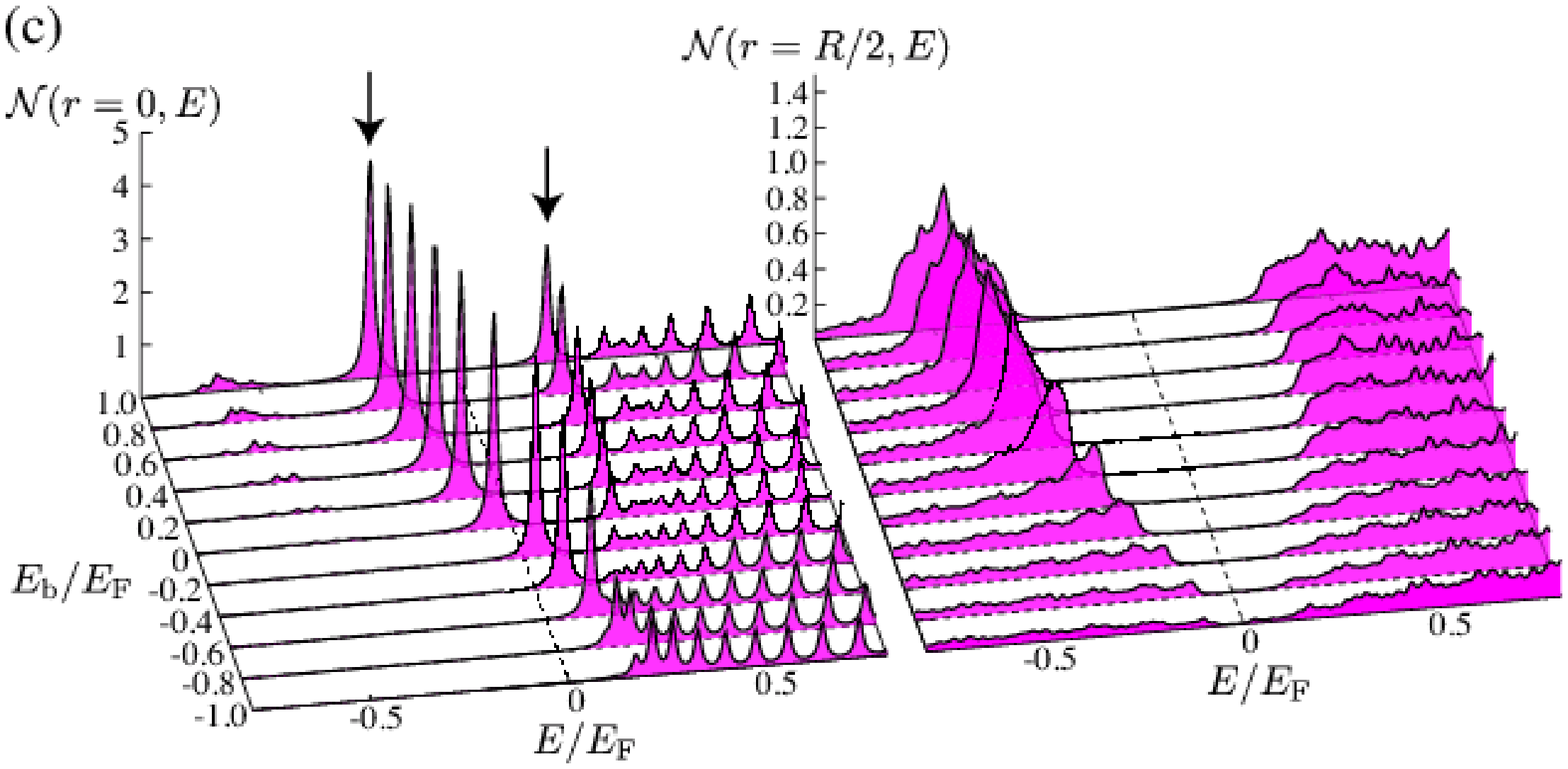}
\caption{(Color online) Local density of states at $r\!=\! 0$ (left column) and $r \!=\! R/2 = 25 k_{\rm F}$ (right column) in the range of $E_{\rm b}/E_{\rm F} \!\in\! [-1.0,1.0]$: (a) $\kappa  \!=\! +1$, (b) $\kappa  \!=\! +2$, and (c) $\kappa  \!=\! +3$. The arrows in the panels denote the CdGM states with $\ell \!=\! 0$. For the LDOS in the case of $\kappa  \!=\! -1$, see Ref.~\cite{mizushima}.
}
\label{fig:ldos}
\end{figure}

To further clarify the relation between $\rho(r\!=\!0)$ and the low-lying quasiparticle state, we present in Fig.~\ref{fig:ldos} the local density of states (LDOS) at the vortex core $\mathcal{N}(r \!=\! 0,E)$ and at the bulk $\mathcal{N}(r \!=\! \frac{R}{2},E)$ in various vortex states, where the LDOS is defined as
\beq
\mathcal{N}({\bm r},E) = \sum _{\nu} |u_{\nu}({\bm r})|^2 \delta _{\eta}(E-E_{\nu}),
\eeq
with the Lorentzian function $\delta _{\eta}(z) \!=\! (\eta/2)^2/[z^2+(\eta/2)^2]$ (see for instance, Ref.~\cite{mizushimaJPS}). In Fig.~\ref{fig:ldos}, we set $\eta \!=\! 0.02 E_{\rm F}$. 

In the case of $\kappa  \!=\! +1$, as we have expected, it is seen in Fig.~\ref{fig:ldos}(a) that $\mathcal{N}(r \!=\! 0,E)$ has the sharp peak about $E/E_{\rm F} \!\approx\! 0.2$ at $E_{\rm b}/E_{\rm F} \!=\! 1.0$ denoted by the arrow in Fig.~\ref{fig:ldos}(a). This peak reflects the lowest CdGM state with $\ell \!=\! 0$, which is not occupied and leads to $\rho _{\rm CdGM}(r \!=\! 0) \!=\! 0$. As the strong coupling regime is approached, the intensity of the peak around $E/E_{\rm F} \!\approx\! 0.2$ is suppressed and merges to the other bulk excitations. In addition, $\rho _{\rm bulk}(r\!=\! 0)$ is constructed from the eigenstates around $E/E_{\rm F}\!\approx\! -0.5$. The negative energy states gradually becomes empty in the vicinity of the BCS-BEC evolution ($E_{\rm b} /E_{\rm F} \!=\! -1.0$), leading to $\rho (r \!=\! 0) \!=\! 0$. In the other vortex states with $\kappa \!=\! +2$ and $+3$, the peak originating from the $\ell \!=\! 0$ CdGM state shifts to the positive energy region at $E_{\rm b}/E_{\rm F} \!\approx\! 0.5$ and $-0.1$, respectively. This results in the abrupt jump of the particle density depletion $\rho(r\!=\!0)$ in the inset of Fig.~\ref{fig:dns}.

\section{Zero energy Majorana states}

It is found from Eqs.~(\ref{eq:edge}) and (\ref{eq:cdgm}) that the energy of the lowest CdGM and edge states can be zero when $\kappa $ is odd and the weak coupling limit $\Delta \!\ll\! E_{\rm F}$ is approached. In contrast, the low-energy spectrum yields an isotropic gap uniquely determined by $|\mu|$ in the BEC phase with $\mu \!<\! 0$. 

\begin{figure}[b!]
\includegraphics[width=80mm]{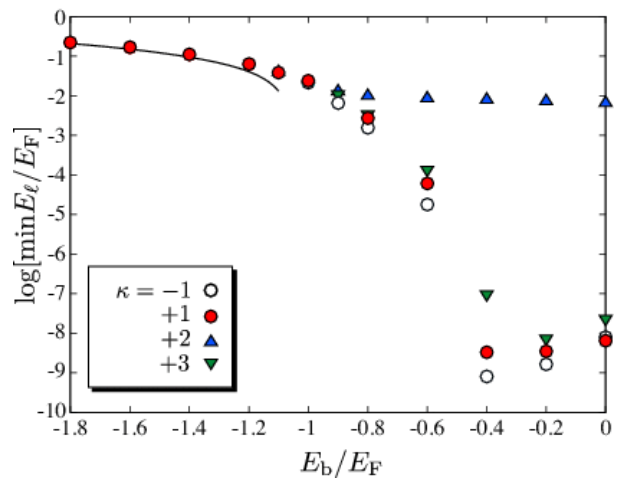}
\caption{(Color online) Lowest eigenenergies with logarithmic scale as a function of $E_{\bm b}/E_{\rm F} \!\in\! [-1.8,0]$ in vortex states with $\kappa  \!=\! -1$ (open circles), $+1$ (filled circles), $+2$ (triangles), and $+3$ (squares). The solid line corresponds to $|\mu|$. 
}
\label{fig:egns}
\end{figure}

Figure~\ref{fig:egns} shows the lowest eigenenergies in the vicinity of the BCS-BEC evolution $E_{\rm b}/E_{\rm F} \!\in\! [-1.8,0.0]$. In the BEC regime $E_{\rm b}/E_{\rm F} \!=\! -1.8$, all the vortex states have the energy gap comparable to $|\mu| \!=\! 0.21 E_{\rm F}$, regardless of $\kappa $. In the case of $\kappa  \!=\! 2$, the lowest eigenenergies stay around $E/E_{\rm F} \!\approx\! 0.01$ even if the weak coupling BCS regime is approached beyond the transition point $E_{\rm b}/E_{\rm F} \!=\! -1.0$. These turn out to be the lowest edge state with $\ell \!=\! +1$, whose energy is in good agreement with the analytic result in Eq.~(\ref{eq:edge}), $E_{\ell \!=\! +1}\!=\!\Delta /(2k_{\rm F}R) \!\approx\! 0.01E_{\rm F}$ with $R \!=\! 50k^{-1}_{\rm F}$ and $\Delta  \!\approx\! E_{\rm F}$.

In contrast to the even $\kappa $ case, the eigenenergies exponentially shift to zero when $\kappa $ is odd. These states are in a consequence of the quasiparticle tunneling between the edge and CdGM states with $\ell \!=\! (\kappa +1)/2 \!\in\! \mathbb{Z}$. Assuming the limit of $R \!\rightarrow \! \infty$, it is known that each state can have the exactly zero energy, as seen in Eqs.~(\ref{eq:edge}) and (\ref{eq:cdgm}) with $\ell \!=\! (\kappa +1)/2 \!\in\! \mathbb{Z}$ and Refs~\cite{tewari,gurarie}. It is known that the zero energy eigenstates exhibit the Majorana property, such as $\Gamma^{\dag}_{E\!=\!0 } \!=\! \Gamma _{E\!=\! 0}$, resulting from $ u_{E=0}({\bm r}) \!=\! v^{\ast}_{E=0}({\bm r})$ \cite{read,ivanov,stone,stone2,gurarie}.

As a consequence of finiteness of the system with $k_{\rm F}R \!=\! 50$, however, these two Majorana wave functions bound at the core $u^{\rm c}_{\ell}({\bm r})$ and at the edge $u^{\rm e}_{\ell}({\bm r})$ are hybridized with each other, resulting in the symmetric and anti-symmetric states, $u^{\rm s}_{\ell} \!=\! [u^{\rm c}_{\ell} + u^{\rm e}_{\ell}]/\sqrt{2}$ and $u^{\rm a}_{\ell} \!=\! [u^{\rm c}_{\ell} - u^{\rm e}_{\ell}]/\sqrt{2}$. The hybridization leads to the splitting of the degenerate zero energy states to the finite energy $\pm E$. Based on the argument about the symmetry of the eigenstates in Sec.~II B, if the wave function of the state with $+E _{\ell}$ is $[u_{\ell},v_{\ell}] \!=\! [u^{\rm s}_{\ell},u^{\rm a}_{\ell}]$, then one finds that the other is $[u^{{\rm a}\ast}_{\ell},u^{{\rm s}\ast}_{\ell}]$ for $-E_{\ell}$.

\begin{figure}[b!]
\includegraphics[width=80mm]{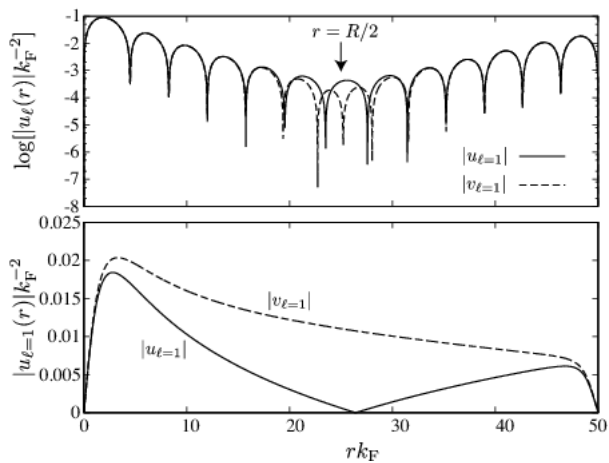}
\caption{Wave functions $|u_{\ell}(r)|$ (solid curve) and $|v_{\ell}(r)|$ (dashed curve) of the CdGM state with $\ell \!=\! 0$ at $E_b/E_F \!=\! 2.0$ (a) and $-1.0$ (b). The eigenenergies are $E/E_{\rm F} \!=\! -1.9\!\times\! 10^{-5}$ in (a) and $E/E_{\rm F} \!=\! -0.024$ in (b).
}
\label{fig:uv}
\end{figure}

Figure~\ref{fig:uv}(a) shows the wave function $|u_{\ell \!=\! 1}(r)|$ and $|v_{\ell \!=\! 1}(r)|$ of the lowest energy state with $E_{\ell \!=\! 1}/E_{\rm F} \!=\! -1.9\!\times \! 10^{-5}$ in the BCS regime of the $\kappa  \!=\! +1$ vortex state. Here, the eigenfunction of this state turns out to be $[u_{\ell},v_{\ell}] \!=\! [u^{\rm s}_{\ell},u^{\rm a}_{\ell}]$. It is seen that the wave function is exponentially localized around the core ($r\!=\! 0$) and around the edge ($r\!=\! 50k^{-1}_{\rm F}$) within the coherence length $\xi \!=\! 2k_{\rm F}/(M\Delta ) \!=\! 4.8k^{-1}_{\rm F}$ with $\Delta  \!=\! 0.42 E_{\rm F}$. Note that due to $\ell \!\neq\! 0$, $|u_{\ell}(r)| \!=\! |v_{\ell}(r)| \!\approx\! r$ is obtained at $r \!\rightarrow\! 0$. The period of rapid oscillation is found to be an order of the $2\pi/k_{\rm \mu} \!=\! 7.2k^{-1}_{\rm F}$ with $\mu \!=\! 0.757 E_{\rm F}$, resulting from $J_{\ell}(k_{\mu}r)$ in Eq.~(\ref{eq:cdgmwf}).

These results displayed in Figs.~\ref{fig:egns} and \ref{fig:uv} are consistent with the analytic expression in two-vortex systems \cite{sarma} and the numerical results in vortex-antivortex systems \cite{kraus,kraus2} and in plural-vortex systems \cite{mizushima09}. In addition, the zero energy states are found to exist in chiral $p$-wave systems within tight binding approximation \cite{takigawa,massignan}. It was proposed in Ref.~\cite{sarma} that due to the rapid oscillation in the wave functions, the {\it Friedel}-like oscillation of the splitting energies appears. This is numerically demonstrated in Ref.~\cite{mizushima09}. Note that the splitting of the zero energy Majorana states, which is critical to the decoherence in the topological quantum computation, is also observed in other systems, such as the non-abelian quasiholes of the $\nu \!=\! \frac{5}{2}$ fractional quantum Hall state \cite{simon,baraban}, Kitaev's honeycomb lattice model \cite{lahtinen}, and the generic anyon model \cite{bonderson}.

As $E_{\rm b}$ decreases from $E_{\rm F}$ to the negative value, the chemical potential touches to the zero, {\it e.g.}, $\mu \!=\! 0.01E_{\rm F}$ at $E_{\rm b}/E_{\rm F} \!=\! -1.0$. Then, the length scale of the wave functions $u_{\ell}$ and $v_{\ell}$ spreads over the whole system, {\it e.g.}, $2\pi/k_{\mu} \!=\! 65 k^{-1}_{\rm F}\!>\! R$ at $E_{\rm b}/E_{\rm F} \!=\! -1.0$, as seen in Fig.~\ref{fig:uv}(b). This also reflects the topological phase transition at $E_{\rm b}/E_{\rm F} \!\approx\! -1.0$ with $\mu \!=\! 0$, where the bulk excitation becomes gapless \cite{read}.

\begin{figure}[b!]
\includegraphics[width=80mm]{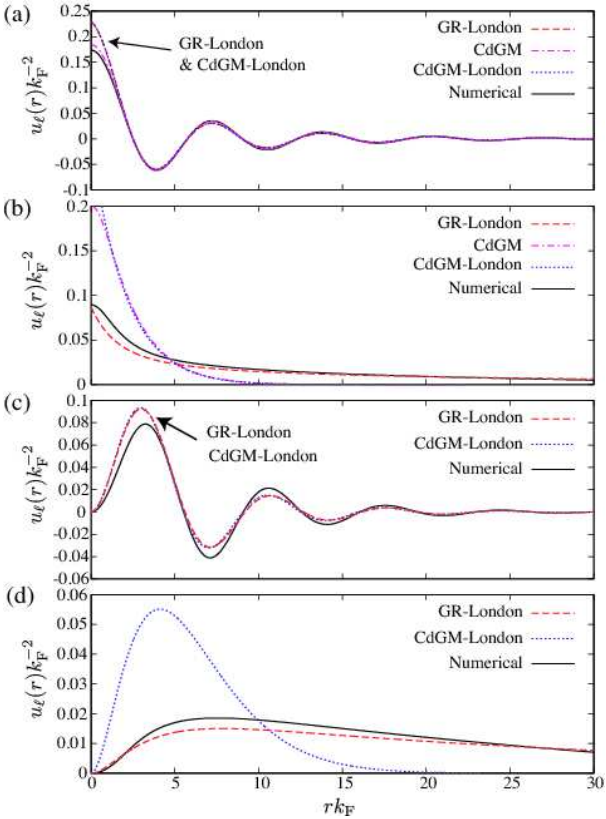}
\caption{(Color online) (a,b) Wave functions $|u_{\ell}(r)|$ with $\ell\!=\!0$ obtained from the full numerical calculation (solid curves) in the $\kappa _{+1} \!=\! -1$ vortex: (a) $E_{\rm b}/E_{\rm F} \!=\! 4.0$ with $k_{\mu}\xi \!=\! 8.6$ and (a) $E_{\rm b}/E_{\rm F} \!=\! -1.0$ with $k_{\mu}\xi \!=\! 0.23$. These are compared with the GR (dashed lines) and CdGM (dotted curves) solutions with the London approximation $\Delta (r) \!=\! \Delta $. The GR and CdGM curves are obtained from Eqs.~(\ref{eq:gr}) and (\ref{eq:cdgmwf}), respectively. In addition, we plot the CdGM solution obtained from Eq.~(\ref{eq:cdgmwf}) (dashed-dotted curves), where $\Delta (r) \!=\! \Delta  \tanh(r/\xi _{\rm core})$ with $\xi _{\rm core} \!=\! 2k^{-1}_{\rm F}$ is assumed. (c,d) Wave functions $|u_{\ell}(r)|$ with $\ell\!=\!2$ obtained from the full numerical calculation (solid curves) in the $\kappa _{+1} \!=\! +3$ vortex: (c) $E_{\rm b}/E_{\rm F} \!=\! 4.0$ with $k_{\mu}\xi \!=\! 8.6$ and (d) $E_{\rm b}/E_{\rm F} \!=\! -1.0$ with $k_{\mu}\xi \!=\! 0.23$. 
}
\label{fig:core}
\end{figure}

Finally, in Fig.~\ref{fig:core}, we compare our numerical results for the wave functions $|u_{\ell}|$ of the lowest CdGM branch of $\kappa \!=\! +1$ and $+3$ vortices with the analytic expressions described in Eq.~(\ref{eq:cdgmwf}) and in Refs.~\cite{tewari07,gurarie}. As described in Appendix B, the CdGM solution assumes the weak coupling limit $k_{\rm F} \xi \!\gg\! 1$, {\it i.e.}, $\Delta \!\ll\! E_{\rm F}$, resulting from the linearization of a fermion dispersion on the Fermi surface. The spatial shape of the resulting wave function consists of two different length scales, such as $J_{\ell}(k_{\mu}r)$ and $e^{-\frac{M}{k_{\rm F}}\int^{r}_0 \Delta (r')dr'}$, as seen in Eq.~(\ref{eq:cdgmwf2}). The former describes the shorter length scale with $k^{-1}_{\mu} \!\approx\! k^{-1}_{\rm F}$ and the latter is slow function varying over $\xi \!\gg\! k^{-1}_{\rm F}$. The alternative solution for the zero energy solution of the $\kappa\!=\! -1$ vortex was discussed by Tewari {\it et al.} \cite{tewari07} in both the BCS and BEC regimes. Gurarie and Radzihovsky (GR) \cite{gurarie} expanded this argument into  the vicinity of the BCS-BEC evolution with arbitrary winding number $\kappa $, who finds that $u_{\ell}$ changes from the Bessel function $J_{\ell}$ to the modified Bessel function $I_{\ell}$ in the strong coupling regime. Following the procedure in Ref.~\cite{gurarie}, the zero energy solution of the BdG equation (\ref{eq:bdgeq}) with $\ell \!=\! (\kappa +1)/2 \!\in\! \mathbb{Z}$ and $E_{\ell}\!=\! 0$ exhibits 
\begin{subequations}
\label{eq:gr}
\beq
u_{\ell}(r) = v_{\ell}(r) = \mathcal{N}J_{\ell}\left(\frac{r}{\xi}\sqrt{(k_{\mu}\xi)^2-1} \right)e^{-r/\xi},
\eeq 
for $k_{\mu}\xi \!>\! 1$ and 
\beq
u_{\ell}(r) = v_{\ell}(r) = \mathcal{N}I_{\ell}\left(\frac{r}{\xi}\sqrt{1-(k_{\mu}\xi)^2} \right)e^{-r/\xi},
\eeq 
\end{subequations}
for $k_{\mu}\xi \!<\! 1$. Here, $I_{\ell}(z)$ is the modified Bessel function and $\mathcal{N}$ is the normalization constant. In Eq.~(\ref{eq:gr}), the London approximation is carried out as $\Delta _{+1} ({\bm r}) \!=\! \Delta e^{i\kappa\theta}$ and $\Delta _{-1}({\bm r}) \!=\! 0$. It is obvious that the GR solution in Eq.~(\ref{eq:gr}) coincides with the CdGM solution in Eq.~(\ref{eq:cdgmwf}) in the BCS limit that $k_{\mu}\xi \!\gg\! 1$.

It is seen from Figs.~\ref{fig:core}(a) and \ref{fig:core}(c) that both the CdGM and GR solutions within the London approximation are in good agreement with the wave functions obtained from the full numerical calculation, at the BCS regime $E_{\rm b}/E_{\rm F} \!=\! 4.0$ with $k_{\mu}\xi \!=\! 8.6$. Here, the deviation around the core can be improved by replacing the constant pair potential to $\Delta _{+1}(r) \!=\! \Delta  \tanh(r/\xi _{\rm c})$, where $\xi _{\rm c} \!=\! \xi/2$ is approximately obtained by fitting with the self-consistent pair potential, {\it c.f.}, see the inset of Fig.~\ref{fig:delta}(a). In the vicinity of the BCS-BEC topological transition, {\it e.g.}, $E_{\rm b}/E_{\rm F}\!=\! -1.0$ with $k_{\mu}\xi \!=\! 0.23$ in Figs.~\ref{fig:core}(b) and \ref{fig:core}(d), the GR solution in Eq.~(\ref{eq:gr}) reproduces the numerical result of $|u_{\ell}(r)|$ in contrast to the CdGM solution. In this regime, the spatial shape of the wave function changes from $J_{\ell}$ to the modified Bessel function $I_{\ell}(r\sqrt{\xi^{-2}-k^2_{\mu}})$.

\section{Concluding remarks}

Here, we have investigated the vortex structures in BCS-BEC evolution of $p$-wave resonant Fermi gases. By using the fully microscopic theory based on the BdG equation, we have revealed in Sec.~III the direct relation between macroscopic vortex structures and low-energy quasiparticle spectra in various vortex states. The low-energy spectrum for a single vortex state with vorticity $\kappa$ is found to consist of the $\kappa$ branches of the CdGM states in addition to the chiral edge state. In particular, it is found that in the strong coupling regime, the quantum depletion appears in the particle density around the vortex core, which reflects the shift of the lowest CdGM state against the effective pairing interaction. Hence, it is proposed that the absorption images around vortex cores can provide the information on the low-lying CdGM state.

In addition, in Sec.~III, we have observed in the $k_x+ik_y$ pairing state with vorticity $\kappa$ that the net angular momentum is well estimated as $\langle L_z \rangle /\hbar \!\approx\! (\kappa  + 1)N/2$ in both the BCS and BEC regimes. This turns out to be distinct from that in the $s$-wave case, which can be estimated as $\langle L_z \rangle /\hbar \!\approx\! \kappa  N/2$. Since this deviation originates from the orbital angular momentum of the chiral $k_x + ik_y$ pairing, the detection of this deviation can be the direct evidence for $p$-wave superfluidity in experiments \cite{mizushima}.

While we assume in this paper a rigid boundary condition at a certain radius without a harmonic trap potential, the Fermi atoms in a realistic system are confined by a three-dimensional harmonic trap which may alter the conclusions that we present here. However, it is inferred that such a harmonic trap does not alter the main outcome obtained here, as long as the single vortex state is assumed. For the CdGM states, this is because the typical size $\xi \!\approx\! k^{-1}_{\rm F}$ of the vortex core is much smaller than the size of the cloud, of which the Thomas-Fermi radius for non-interacting systems is estimated through $R_{\rm TF} \!=\! \sqrt{2E_{\rm F}/M\omega^2}$ approximately as an order of 10 $\mu {\rm m}$, where $\omega$ is the trap frequency. Hence, the curvature due to the trap potential may be much smaller than the spatial variation of the order parameter around the vortex and is ineffective to the results on the CdGM states and the depletion of the particle density. It has also been demonstrated in our previous work~\cite{mizushima} that the trap potential does not alter the low energy dispersion of the edge states in the vortex-free state as well as the net angular momentum $\langle L_z \rangle /\hbar \!\approx\! N/2$. However, such a potential might work against the stability of the Majorana zero modes when the cloud contains many vortices under rapid rotation.

Furthermore, it has been discussed in Sec.~IV that the zero energy state appears in the BCS phase of an odd $\kappa$ vortex state. It is revealed that the lowest eigenenergies exponentially lift from zero to finite values as the strong coupling regime is approached. This is because the length scale of the wave function in the CdGM and edge state with $E_{\ell} \!=\! 0$ becomes comparable to the macroscopic system size and the CdGM and edge states having a degenerate energy are hybridized with each other. In the extensive range from the weak coupling BCS regime to the BCS-BEC topological phase transition point, the wave function of the zero energy state is compared with the analytic expressions, such as the CdGM solution in Eq.~(\ref{eq:cdgmwf}) and the GR solution in Ref.~\cite{gurarie}. Here, it is demonstrated that the spatial profile turns out to exhibit not $J_{\ell}$ but the modified Bessel function $I_{\ell}$ in the strong coupling regime with $k_{\mu}\xi \!<\! 1$ \cite{gurarie}. 

Finally, it is worth mentioning that the zero energy states are unchanged by a slow rotation of the system. This can be demonstrated that for axisymmetric vortices with arbitrary winding numbers $\langle \kappa _{+1}, \kappa _{0}, \kappa _{-1}\rangle \!=\! \langle \kappa , \kappa +1, \kappa +2\rangle$, the eigenenergy $E_{\ell}(\Omega)$ obtained from the BdG equation (\ref{eq:bdgeq}) under rotation with a frequency $\Omega$ yields a linear shift from the value without rotation $E_{\ell}(\Omega \!=\! 0)$ as 
\begin{eqnarray}
E_{\ell}(\Omega) = E_{\ell}(\Omega \!=\! 0) - \left(\ell - \frac{\kappa  +1}{2} \right)\Omega,
\label{eq:omega} 
\end{eqnarray}
under slow rotation within $\Omega \!\ll\! E_F$. This implies that the zero energy state with $\ell \!=\! (\kappa +1)/2\!\in\! \mathbb{Z}$ remains as it is, regardless of $\Omega$. In this argument, we neglect the possibility on the change of the pair potentials $\Delta _{\pm 1}({\bm r})$, such as the penetration of the more vortices. The self-consistent pairing field and the existence of the zero energy state in a rapidly rotating regime is open for future study.

\section*{ACKNOWLEDGMENTS}


The authors are grateful to L. Radzihovsky for various insightful discussions and fruitful comments to the manuscript. This research was supported by the Grant-in-Aid for Scientific Research, Japan Society for the Promotion of Science.

\appendix

\section{Self-consistent equations}

Here, we describe the details on the derivation of the BdG equation (\ref{eq:bdgeq}) and gap equation (\ref{eq:gapeq}). First, in order to diagonalize the mean-field Hamiltonian in Eq.~(\ref{eq:Hmf}), we introduce the unitary transformation to the quasiparticle basis with ${\bm \eta}_{\nu} \!\equiv\! [\eta _{\nu}, \eta^{\dag}_{\nu}]^T$, 
\begin{eqnarray}
{\bm \Psi}({\bm r}_1) = \sum _{\nu} \hat{u}_{\nu}({\bm r}_1) {\bm \eta}_{\nu}, 
\label{eq:bogolon}
\end{eqnarray}
where $\hat{u}_{\nu}$ is a $2\!\times\!2$ matrix and the matrix elements describe the quasiparticle wave functions. They are required to satisfy the orthonormal condition,
\begin{eqnarray}
\int \hat{u}^{\dag}_{\nu}({\bm r}_1)\hat{u}_{{\nu}'}({\bm r}_1) d{\bm r}_1 \!=\! \delta _{\nu,\nu'}, 
\label{eq:ortho}
\end{eqnarray}
and completeness conditions,
$\sum _{\nu} \hat{u}_{\nu}({\bm r}_1)\hat{u}^{\dag}_{\nu}({\bm r}_2) \!=\! \delta({\bm r}_1-{\bm r}_2)$. Also, the fermion operators $\eta _{\nu}$ and $\eta^{\dag}_{\nu}$ obey the anti-commutation relation, $\{ \eta _{\nu}, \eta^{\dag}_{{\nu}'} \} \!=\! \delta _{{\nu},{\nu}'}$, $\{ \eta _{\nu}, \eta _{{\nu}'} \} \!=\! \{ \eta^{\dag}_{\nu}, \eta^{\dag}_{{\nu}'} \} \!=\! 0$. The mean-field Hamiltonian in Eq.~(\ref{eq:Hmf}) is now diagonalized in terms of this basis with the quasiparticle energy $E_{\nu}$ as
$\mathcal{H} \!=\! E_0 + \frac{1}{2}\sum E_{\nu} {\bm \eta}^{\dag}_{\nu} \hat{\tau}_3 {\bm \eta}_{\nu}$.
This diagonalization leads to the Bogoliubov-de Gennes (BdG) equation, 
$\int d{\rm r}_2 \hat{\mathcal{K}}({\bm r}_1,{\bm r}_2) \hat{u}_{\nu}({\bm r}_2) \!=\! E_{\nu} \hat{u}_{\nu}({\bm r}_1) $. Here, it is found that the matrix elements of $\hat{u}$ yield $(\hat{u})_{22} \!=\! (\hat{u})^{\ast}_{11}$ and $(\hat{u})_{12} \!=\! (\hat{u})^{\ast}_{21}$, because of the symmetry,
$
-\hat{\tau}_1\hat{\mathcal{K}^{\ast}}({\bm r}_1,{\bm r}_2)\hat{\tau}_1 \!=\! \hat{\mathcal{K}}({\bm r}_1,{\bm r}_2)
$,
where $\hat{\tau}_{1,2,3}$ denote the Pauli matrices. Hence, the BdG equation reduces to the equation for the quasiparticle wave functions $u_{\nu} \!=\! (\hat{u})_{11}$ and $v_{\nu} \!=\! (\hat{u})_{21}$,
\begin{eqnarray}
\int d{\bm r}_2 \hat{\mathcal{K}}({\bm r}_1,{\bm r}_2)
\left[
\begin{array}{c} u_{\nu}({\bm r}_2) \\ v_{\nu}({\bm r}_2) \end{array}
\right]
=  E_{\nu}
\left[
\begin{array}{c} u_{\nu}({\bm r}_1) \\ v_{\nu}({\bm r}_1) \end{array}
\right].
\label{eq:bdg_general}
\end{eqnarray}

We then expand the pair potential $\Delta ({\bm r}_1,{\bm r}_2)$ in Eq.~(\ref{eq:delta}) to the Fourier series with respect to the relative coordinate ${\bm r}_{12} \!\equiv\! {\bm r}_1 - {\bm r}_2$ as, 
\beq
\Delta ({\bm r}_1,{\bm r}_2) \!=\! \int \frac{d{\bm k}}{(2\pi)^3} e^{i{\bm k}\cdot{\bm r}_{12}}
\Delta({\bm r},{\bm k}) ,
\label{eq:deltaexp}
\eeq
with the relative wave vector ${\bm k}$.The Fourier coefficient $\Delta({\bm r},{\bm k})$ is assumed to be expanded in terms of a $p$-wave channel,
\beq
\Delta ({\bm r},{\bm k}) = \Gamma (k)  \sum _{m = 0, \pm 1} \Delta _m({\bm r}) \hat{k}_m.
\label{eq:gap_formal}
\eeq
where we set $\Gamma (k) \!=\! k/k_0$ and $\hat{k}_{\pm 1} \!\equiv\! \mp i (\hat{k}_x\pm i\hat{k}_y)$ and $\hat{k}_0 \!\equiv\! -i\hat{k}_z$. 
We will later see that the pair potential $\Delta _m({\bm r})$ in the $m$ orbital channel can be expressed in terms of the Bogoliubov quasiparticle with the wave function $(u_{\nu},v_{\nu})$ and the energy $E_{\nu}$.

By substituting Eq.~(\ref{eq:gap_formal}) into Eq.~(\ref{eq:deltaexp}) and following the procedure in Ref.~\cite{matsumoto2}, the off-diagonal element in the BdG equation (\ref{eq:bdg_general}) is rewritten as  
\beq
\Delta ({\bm r}_1,{\bm r}_2)
= \frac{1}{k_0} \sum _{m}
  \Delta _m ({\bm r}) \mathcal{P}^{(1)}_{m}\delta({\bm r}_1-{\bm r}_2).
\eeq
Hence, the BdG equation for the quasiparticles under pair potentials in $m$ orbital channels are now obtained by
\begin{subequations}
\label{eq:bdg_fin}
\beq
\left[
\begin{array}{cc}
H_0({\bm r}) & \Pi ({\bm r}) \\ -\Pi^{\ast}({\bm r}) & -H^{\ast}_0({\bm r})
\end{array}
\right]\left[
\begin{array}{c}
u_{\nu}({\bm r}) \\ v_{\nu}({\bm r})
\end{array}
\right] = E_{\nu}
\left[
\begin{array}{c}
u_{\nu}({\bm r}) \\ v_{\nu}({\bm r})
\end{array}
\right] 
\eeq
\beq
\Pi ({\bm r}) = \frac{1}{k_0}\sum _{m = 0, \pm 1}
\bigg[
\Delta _{m}({\bm r})\mathcal{P}_m + \frac{1}{2}\mathcal{P}_m\Delta _m({\bm r})
\bigg],
\eeq
\end{subequations}
where the $p$-wave operator $\mathcal{P}_m$ is defined as
$\mathcal{P}_{\pm 1} \!\equiv\! \mp  \left( \partial _x \pm i \partial _y \right)$ and
$\mathcal{P}_0 \!\equiv\! \partial _z$. 
The wave functions must satisfy the orthonormal condition in Eq.~(\ref{eq:ortho})
\beq
\int \left[ u^{\ast}_{\nu}({\bm r}) u_{\mu}({\bm r})
+ v^{\ast}_{\nu}({\bm r})v_{\mu}({\bm r})\right]d{\bm r}
= \delta _{\nu,\mu}.
\label{eq:normal3}
\eeq

Then, in order to derive the gap equation for $\Delta _m({\bm r})$, we express the Fourier coefficient $\Delta({\bm r},{\bm k})$ in Eq.~(\ref{eq:deltaexp}) with Eqs.~(\ref{eq:delta}) and (\ref{eq:bogolon}) as
\begin{subequations}
\label{eq:gap1}
\beq
\Delta({\bm r},{\bm k}) 
= \int d{\bm r}_{12}e^{-i{\bm k}\cdot{\bm r}_{12}}V(\tilde{r})\Phi({\bm r}_1,{\bm r}_2)
\eeq
\beq
\Phi({\bm r}_1,{\bm r}_2) \equiv
\sum _{\nu} v^{\ast}_{\nu}({\bm r}_2) u_{\nu}({\bm r}_1) f(E_{\nu}).
\label{eq:phi}
\eeq 
\end{subequations}
Here, we assume the following $p$-wave symmetric interparticle interaction, 
\beq
V_{\bm k} \equiv \int d{\bm r}_{12} e^{-i{\bm k}\cdot{\bm r}_{12}}V(\tilde{r})
= \sum _{m\!=\! 0, \pm 1} g_m \left| \Gamma (k) \hat{k}_{m} \right|^2,
\label{eq:vk} 
\eeq
where $g_m \!<\! 0$ is the coupling constant for the $m$ orbital channel of the scattering. Assuming the limit of the ${\bm r}_{12} \!\rightarrow\! 0$, the Taylor expansion of the pair function $\Phi$ gives 
\beq
&& \hspace{-10mm} \Phi ({\bm r}_1,{\bm r}_2) \approx \Phi({\bm r},{\bm r}) \nn \\ 
&& + \bigg[ 
(\nabla _1 - \nabla _2)\Phi({\bm r}_1,{\bm r}_2) 
\bigg]_{{\bm r}_{12} \rightarrow 0} \cdot \frac{{\bm r}_{12}}{2}.
\label{eq:phi2}
\eeq
Then, by substituting the interaction $V_{\bm k}$ in Eq.~(\ref{eq:vk}) and $\Phi$ in Eq.~(\ref{eq:phi2}) into the gap equation (\ref{eq:gap1}), one finally finds
\beq
&& \hspace{-5mm} \Delta ({\bm r},{\bm k}) =
\bigg[ \sum _m g_m\left| \Gamma (k) Y_{1,m}(\hat{\bm k})\right|^2\bigg] \Phi({\bm r},{\bm r}) \nn \\
&& \hspace{-5mm}+ \frac{g_0}{k_0}\Gamma(k)
\bigg\{
\bigg[ \gamma \bigg(\mathcal{P}^{(1)}_{-1} 
- \mathcal{P}^{(2)}_{-1}\bigg) \Phi({\bm r}_1,{\bm r}_2)\bigg]_{{\bm r}_2 \rightarrow {\bm r}_1} 
\hat{k}_{+1} \nn \\
&& \hspace{-5mm}- \bigg[ \bigg(\mathcal{P}^{(1)}_{0} 
- \mathcal{P}^{(2)}_{0}\bigg) \Phi({\bm r}_1,{\bm r}_2)\bigg]_{{\bm r}_2 \rightarrow {\bm r}_1} 
\hat{k}_{0} \nn \\
&& \hspace{-5mm}+ \bigg[ \gamma \bigg(\mathcal{P}^{(1)}_{+1} 
- \mathcal{P}^{(2)}_{+1}\bigg) \Phi({\bm r}_1,{\bm r}_2)\bigg]_{{\bm r}_2 \rightarrow {\bm r}_1} 
\hat{k}_{-1}
\bigg\}.
\eeq
Here, the first term should be zero due to the $p$-wave interaction. With the expression in Eq.~(\ref{eq:gap_formal}), one reads the gap equations in local form as 
\begin{subequations}
\label{eq:gap}
\beq
\Delta _{\pm 1} ({\bm r}) = \frac{g_{\pm 1}}{k_0}
\bigg(\mathcal{P}^{(1)}_{\mp 1} 
- \mathcal{P}^{(2)}_{\mp 1}\bigg) \Phi({\bm r}_1,{\bm r}_2)\bigg|_{{\bm r}_2 \rightarrow {\bm r}_1} ,
\eeq
\beq
\Delta _{0} ({\bm r}) = -\frac{g_0}{k_0}
\bigg[ \bigg(\mathcal{P}^{(1)}_{0} 
- \mathcal{P}^{(2)}_{0}\bigg) \Phi({\bm r}_1,{\bm r}_2)\bigg|_{{\bm r}_2 \rightarrow {\bm r}_1} ,
\eeq
\end{subequations}
The sum $\sum_{\nu}$ in Eq.~(\ref{eq:gap}) denotes the summation for all the eigenstates with the positive and negative eigenvalues.

\section{Core-bound states in a chiral $p$-wave superfluid}

In this Appendix, we describe the details on the derivation of the analytic expression of the CdGM states in Eq.~(\ref{eq:cdgm}), from the BdG equation (\ref{eq:bdgeq}). Without loss of generality, let us consider the situation that $k_x+ik_y$ pairing state is majority and $k_z$ component is negligible, {\it i.e.}, 
\begin{subequations}
\begin{eqnarray}
\Delta _{+1}({\bm r}) = \Delta _{+1}(r)e^{i\kappa \theta} , 
\eeq
\beq
\Delta _{-1}({\bm r}) = \Delta _{-1}(r)e^{i(\kappa+2) \theta},
\end{eqnarray}
\end{subequations}
and $\Delta _0 ({\bm r}) \!=\! 0$. The CdGM solution has been obtained by Kopnin and Salomaa \cite{kopnin} for the negative vortex state, where the vorticity is anti-parallel to the chirality of the pairing ($\kappa\!=\! -1$), and by Stone and Chung \cite{stone} for the more general case. Also, Volovik \cite{volovik} has analytically solved the BdG equation (\ref{eq:bdgeq}), based on the quasiclassical approximation with a quantization rule. As long as the zero energy states, the alternative expression was derived in Refs.~\cite{tewari07,gurarie}. Here, we expand the expression derived in Refs.~\cite{kopnin,stone} into the more general form which is applicable to vortex systems with an arbitrary winding number $\kappa$ and the induced component $\Delta _{-1}$. The important consequence obtained here is that the qualitative results on the CdGM state are unchanged by the minority $\Delta _{-1}$ component.

We start with the BdG equation (\ref{eq:bdgeq}) in the cylindrical coordinate described in Sec.~II B. Assuming $q \!\equiv\! k_{\mu}\sqrt{1-\sin^2(\alpha)} \!\ll\! k_{\mu} \!\equiv\! \sqrt{2M|\mu|^2}$, the BdG equation (\ref{eq:bdgeq}) reduces to
\begin{eqnarray}
&& \hspace{-5mm}
\left[ \mathcal{L}_m \hat{\tau}_0 
+ \frac{M}{k_{\rm F}}\left\{D_- \frac{d}{dr} + \frac{1}{2} \frac{dD_-}{dr} + \frac{D_-}{2r}
\right\}\hat{\tau}_2  
+ 2M E_{\nu}\hat{\tau}_3 \right] {\bm u}_{\nu} \nn \\
&& + \bigg[ \frac{(\kappa+1)(\ell-\frac{\kappa+1}{2})}{r^2}
 \hat{\tau}_3    -i\frac{MD_+}{k_{\rm F}}\frac{\ell-\frac{\kappa+1}{2}}{ r } \hat{\tau}_1
\bigg]{\bm u}_{\nu},
\label{eq:bdg2}
\end{eqnarray}
where we set $D_{\pm} \!=\! D_{\pm}(r) \!\equiv\! \Delta _{+ 1}(r) \pm \Delta _{-1}(r)$, ${\bm u}_{\nu} \!\equiv\! {\bm u}_{\nu}(r) \!=\! [u_{\nu}(r),v_{\nu}(r)]^T$, and $\nu \!=\! (n,\ell,q)$. Also, we introduce $\mathcal{L}_m \!\equiv\! \frac{d^2}{dr^2} + \frac{1}{r}\frac{d}{dr} - \frac{m^2}{r^2} + k^2_{\mu}\sin^2(\alpha)$ with $m \!=\! \sqrt{\ell^2-(\kappa+1)\ell+\frac{(\kappa+1)^2}{2}}$ and $2\!\times\!2$ unit matrix $\hat{\tau}_0\!=\! {\rm diag}(1,1)$. Throughout this Appendix, we consider the BCS regime with $\mu \!>\! 0$. 

Following the procedure proposed by Caroli {\it et al.} \cite{cdgm}, we introduce a radius $r_c$ that $\Delta(r) \!=\! 0$ for $r\!<\!r_c$. Then, the BdG equation (\ref{eq:bdg2}) can be analytically solved if the following conditions are assumed: (i) $|\ell| \!\ll\! r_c k_{\rm F} \!\ll\! k_{\rm F}\xi$, (ii) $k_{\rm F}\xi \!\gg\! 1$, (iii) $|E_{\nu}| \!\ll\! |\mu|^2\sin^2(\alpha)$, and (iv) $|\Delta _{+1}(r) | \!\gg\! |\Delta _{-1}(r) |$. These conditions restrict to the low-energy states in weak coupling BCS regime. 

The solution in Eq.~(\ref{eq:bdg2}) is obtained in the range of $r \!<\! r_c$ as 
\begin{eqnarray}
{\bm u}_{\nu}(r) = \mathcal{N}
\left[
\begin{array}{c}
J_{\ell}(k_+(\alpha) r) \\ J_{\ell-\kappa-1} (k_-(\alpha)r)
\end{array}
\right],
\label{eq:uv_r0} 
\end{eqnarray} 
where $\mathcal{N}$ is the arbitrary constant and we set
\begin{eqnarray}
k_{\pm}(\alpha) \!\equiv\! k_{\mu}\sin(\alpha)\pm \frac{E_{\nu}}{v_{\mu}(\alpha)},
\label{eq:kpm}
\end{eqnarray}
with $v_{\mu}(\alpha) \!=\! k_{\mu}\sin(\alpha)/M$. 

For $r\!>\! r_c$, the wave functions are assumed to consist of the Hankel function $H^{(i)}_m$ and the slow functions varying over the order of $\xi$, ${\bm \varphi}_i(r)$, as
\begin{eqnarray}
{\bm u}_{\nu}(r) = \sum _{i=1,2} H^{(i)}_m (k_{\mu}\sin(\alpha)r){\bm \varphi}_i(r).
\label{eq:hankel}
\end{eqnarray}
Then, Eq.~(\ref{eq:bdg2}) within the conditions (i)-(iii) reduces to 
\begin{eqnarray}
&& \hspace{-5mm}
\left[ \hat{\tau}_0 \frac{d}{dr} + \hat{\tau}_2\frac{MD_-(r)}{k_{\rm F}} \right]
{\bm \varphi}_1 (r) 
= \bigg[ -\hat{\tau}_1 \frac{D_+(r)}{k_{\rm F}v_{\mu}(\alpha)} \frac{\ell-\frac{\kappa+1}{2}}{r} \nn \\
&& \hspace{5mm}
+ i\hat{\tau}_3 
\left\{ \frac{E_{\nu}}{v_{\mu}(\alpha)} - \frac{(\kappa+1)(\ell-\frac{\kappa+1}{2})}{2Mv_{\mu}(\alpha)r^2} 
\right\} 
\bigg] {\bm \varphi}_1(r) ,
\label{eq:bdg3}
\end{eqnarray}
and ${\bm \varphi}_2 (r) \!\propto\! \hat{\tau}_3{\bm \varphi}^{\ast}_1 (r)$. Under the conditions (i)-(iii) described above, the right hand side of Eq.~(\ref{eq:bdg3}) can be regarded as the small perturbation. Hence, the regular solution within the first order on $\psi _{1,2}$ is given by
\begin{eqnarray}
{\bm \varphi}_1 (r) = {\bm \varphi}^{(0)}_1 (r) + iB_1 e^{-\chi(r)} 
\left[ \begin{array}{c} \psi _1 (r) \\ i\psi _2 (r) \end{array} \right],
\label{eq:egnvec}
\end{eqnarray}
where ${\bm \varphi}^{(0)}_1 (r)$ is the regular solution when the right hand side of Eq.~(\ref{eq:bdg3}) is neglected, {\it i.e.}, 
${\bm \varphi}^{(0)}_1 (r) = B_1 e^{-\chi(r)} [ 1, i]^{\rm T}$, and 
\begin{eqnarray}
\chi(r) \equiv \frac{M}{k_{\rm F}} \int^{r}_0 \left[ \Delta _{+1} (r')-\Delta _{-1}(r')\right] dr'.
\end{eqnarray}
Since $|\psi _{1,2}(r)| \!\ll\! 1$, Eq.~(\ref{eq:egnvec}) can be also expressed as
\begin{eqnarray}
{\bm \varphi}_1 (r) \simeq B_1e^{-\chi(r)} 
\left[ \begin{array}{c} e^{i\psi _1 (r)} \\ ie^{i\psi _2 (r)} \end{array} \right].
\label{eq:egnvec2}
\end{eqnarray}
Within Eq.~(\ref{eq:egnvec}), one can find the solution of Eq.~(\ref{eq:bdg3}) as $\psi _1 (r) \!=\! - \psi _2 (r) \!\equiv\! \psi(r)$, where
\begin{eqnarray}
&& \hspace{-10mm}
\psi (r)  = - \int^{\infty}_r 
\bigg[ \frac{E_{\nu}}{v_{\mu}(\alpha)} 
- \frac{(\kappa+1)(\ell-\frac{\kappa+1}{2})}{2Mv_{\mu}(\alpha){r'}^{2}} \nn \\
&& 
- \frac{D_+(r')}{v_{\rm F}} \frac{\ell-\frac{\kappa+1}{2}}{Mv_{\mu}(\alpha)r'}
\bigg]
e^{-2\{(\chi(r') -\chi(r)\}}dr'.
\label{eq:psi0}
\end{eqnarray}

In order to obtain the solution of the BdG equation (\ref{eq:bdg2}), the wave functions in Eq.~(\ref{eq:uv_r0}) for $r\!<\! r_c$ and Eq.~(\ref{eq:hankel}) for $r\!>\! r_c$, are now matched at $r \!=\! r_c$. Because of the condition (i), $|\ell| \!\ll\! r_c k_{\rm F}$, making use of the asymptotic forms of $J_{\ell}(z)$ and $H^{(1,2)}_{\ell}(z)$ in $z\!\gg\!|\ell|$, the wave functions for $r\!<\! r_c$ in Eq.~(\ref{eq:uv_r0}) is rewritten as
\beq
{\bm u}_{\nu} \approx \mathcal{N}\sqrt{\frac{2M}{\pi v_{\mu}r}}
\left[
\begin{array}{c}
\displaystyle{\cos{
\left( 
k_{+}r + \frac{\ell^2-\frac{1}{4}}{2k_{+}r} - \frac{2\ell+1}{4} \pi 
\right)}} \\
\displaystyle{\cos{
\left( 
k_{-}r + \frac{\ell^2_v-\frac{1}{4}}{2k_{-}r} - \frac{2\ell_v+1}{4} \pi 
\right)}} 
\end{array}
\right], \nn \\
\label{eq:solution1}
\eeq
with $v_{\mu}\!\equiv\!v_{\mu}(\alpha)$ and $k_{\pm}\!\equiv\! k_{\pm}(\alpha)$. Also, Eq.~(\ref{eq:hankel}) with Eq.~(\ref{eq:egnvec2}) for $rk_{\rm F} \!>\! r_ck_{\rm F} \!\gg\! |\ell|$ is 
\begin{eqnarray}
\hspace{-5mm}
{\bm u}_{\nu} \approx
\sqrt{\frac{2M}{\pi v_{\mu}r}} e^{-\chi(r)}
\left[
\begin{array}{c}
\displaystyle{B_1e^{i\eta _{+}(r)} + B_2 e^{-i\eta _+(r)}} \\
\displaystyle{iB_1e^{i\eta _{-}(r)} + i B_2 e^{-i\eta _-(r)}}
\end{array}
\right],
\label{eq:solution2}
\end{eqnarray}
with
\begin{eqnarray}
\eta _{\pm}(r) \equiv Mv_{\mu}r + \frac{m^2 - \frac{1}{4}}{2Mv_{\mu}r} 
- \frac{2m+1}{4}\pi \pm \psi (r) .
\label{eq:eta}
\end{eqnarray}

Then, the coefficients $B_{1,2}$ are determined so as to match two expressions of ${\bm u}_{\nu}(r)$ in Eqs.~(\ref{eq:solution1}) and (\ref{eq:solution2}) at $r\!=\!r_c$ as
\begin{eqnarray}
B_1 = \frac{\mathcal{N}}{2} e^{i\gamma}, \hspace{3mm} B_2 = \frac{\mathcal{N}}{2} e^{-i\gamma}.
\end{eqnarray}
By comparing with Eqs.~(\ref{eq:solution1}) and (\ref{eq:solution2}), one can obtain the expression of $\psi$ as
\begin{eqnarray}
\psi(r) \approx \frac{E_{\nu}}{v_{\mu}(\alpha)}r 
+ \frac{(\kappa+1)(\ell-\frac{\kappa+1}{2})}{2Mv_{\mu}(\alpha)r}  + \frac{\pi}{2}(m-\ell) - \gamma. \nn \\
\label{eq:psi1}
\end{eqnarray}
In the same way, the another expression is obtained from Eqs.~(\ref{eq:solution1}) and (\ref{eq:solution2}) 
\begin{eqnarray}
\psi(r) &\approx& \frac{E_{\nu}}{v_{\mu}(\alpha)}r 
+ \frac{(\kappa+1)(\ell-\frac{\kappa+1}{2})}{2Mv_{\mu}(\alpha)r} \nn \\ 
&& - \frac{\pi}{2}(m-\ell+\kappa+1) + \gamma - n\pi ,
\label{eq:psi2}
\end{eqnarray}
where $n \!\in\! \mathbb{Z}$. The expressions on $\psi(r)$ in Eqs.~(\ref{eq:psi1}) and (\ref{eq:psi2}) becomes identical when $\gamma$ satisfies
\begin{eqnarray}
\gamma = \frac{\pi}{2} \left( m - \ell + \frac{\kappa+1}{2} \right) + \frac{\pi}{2}n. 
\label{eq:gamma}
\end{eqnarray}

The alternative expressions of $\psi(r)$ in Eq.~(\ref{eq:psi0}) and Eq.~(\ref{eq:psi1}) with Eq.~(\ref{eq:gamma}) should be identical at $r \!=\! r_c$. Hence, we finally obtain the eigenvalue of the BdG equation (\ref{eq:bdg_fin}), 
\begin{eqnarray}
\hspace{-5mm}
E_{\nu} = - \left( \ell-\frac{\kappa+1}{2}\right) \omega _0
+ \left( n - \frac{\kappa+1}{2} \right) \sin(\alpha)\omega _1, 
\label{eq:energy}
\end{eqnarray}
where 
\begin{subequations}
\label{eq:analyticE}
\begin{eqnarray}
\omega _0 \equiv 
\frac{\displaystyle{\int^{\infty}_{r_c} \frac{\kappa _{+1}\Delta_{+1}(r')-\kappa _{-1}\Delta _{-1}(r')}{k_{F}r'} e^{-2\chi(r')}dr'}}
     {\displaystyle{\int^{\infty}_0e^{-2\chi(r')}dr'}}, \nn \\
\end{eqnarray}
\begin{eqnarray}
\omega _1 \equiv
\frac{\pi k_{\mu}}{2M \displaystyle{\int^{\infty}_0e^{-2\chi(r')}dr'}}.
\end{eqnarray}
\end{subequations}
The resulting eigenvalue in Eq.~(\ref{eq:energy}) reproduces the results derived in Res.~\cite{kopnin,stone} and the qualitative properties, such as the appearance of the ZES, turn out to be unchanged by the minority component $\Delta _{-1}$. 

The eigenfunction is then obtained from Eqs.~(\ref{eq:solution2})-(\ref{eq:gamma}) with $k_{\pm}(\alpha)$ in Eq.~(\ref{eq:kpm}) as
\beq
&& \hspace{-13mm} \left[
\begin{array}{cc} 
u_{n,\ell,q}(r) \\ v_{n,\ell,q}(r)
\end{array}
\right] = \mathcal{N}
\left[
\begin{array}{cc} 
J_{\ell}(k_{+}(\alpha)r) \\ J_{\ell-\kappa-1}(k_{-}(\alpha)r)
\end{array}
\right] \nn \\
&& \times \exp\left\{-\frac{M}{k_{\rm F}}\int^{r}_0 \left[\Delta _{+1}(r^{\prime})-\Delta _{-1}(r^{\prime})\right]dr^{\prime}\right\}. 
\eeq
To estimate the order of the energy scale of $\omega _{0,1}$, let us consider the simplest case of $\Delta(r)$, that is, $\Delta _{+1}(r) \!=\! \Delta \tanh{(r/\xi)}$ and $\Delta _{-1}(r) \!=\! 0$. In this situation, one finds $\omega _1 \!\approx\! \frac{\pi}{2}\Delta$ and $\omega _0 \!\approx\! \kappa\frac{\Delta^2}{E_F}$ with $\int^{\infty}_0 e^{-2\chi(r')}dr' \!=\! \xi$. Hence, the eigenvalue $E_{\nu}$ consists of two different energy scales, such as $\Delta $ and $\Delta^2/E_F$. The expression in the case of a singly quantized vortex ($\kappa\!=\!\pm 1$) coincides to that in Refs.~\cite{kopnin} and \cite{stone}, which yields the zero energy modes with $\ell\!=\! 0$ for $\kappa\!=\! -1$ and $\ell\!=\! -1$ for $\kappa\!=\! +1$.



\begin{thebibliography}{99}

\bibitem{legett}
A.J. Leggett, Rev. Mod. Phys. {\bf 47}, 331 (1975).

\bibitem{vollhardt}
D. Vollhardt and P. W\"{o}lfle, {\it The Superfluid Phases of Helium 3} (Taylor \& Francis, London, 1990).

\bibitem{salomaa}
M.M. Salomaa and G.E. Volovik, Rev. Mod. Phys. {\bf 59}, 533 (1987).

\bibitem{volovik1}
G.E. Volovik, {\it Universe in a Helium Droplet} (Oxford University Press, 2003).

\bibitem{read}
N. Read and D. Green, Phys. Rev. B {\bf 61}, 10267 (2000).

\bibitem{salomon}
J. Zhang, E.G.M. van Kempen, T. Bourdel, L. Khaykovich, J. Cubizolles, F. Chevy, M. Teichmann, L. Tarruell, S.J.J M.F. Kokkelmans, and C. Salomon, Phys. Rev. A {\bf 70}, 030702(R) (2004).

\bibitem{ketterle}
C H. Schunck, M.W. Zwierlein, C.A. Stan, S.M.F. Raupach, and W. Ketterle, A.Simoni, E.Tiesinga, C.J. Williams, and P.S. Julienne, 
Phys. Rev. A {\bf 71}, 045601 (2005).

\bibitem{jin}
J.P. Gaebler, J.T. Stewart, J.L. Bohn, and D.S. Jin, Phys. Rev. Lett. {\bf 98}, 200403 (2007).

\bibitem{fuchs}
J. Fuchs, C. Ticknor, P. Dyke, G. Veeravalli, E. Kuhnle, W. Rowlands, P. Hannaford, and C.J. Vale,
Phys. Rev. A {\bf 77}, 053616 (2008).

\bibitem{inada}
Y. Inada, M. Horikoshi, S. Nakajima, M. Kuwata-Gonokami, M. Ueda, and T. Mukaiyama,
Phys. Rev. Lett. {\bf 101}, 100401 (2008).

\bibitem{crossover}
A. J. Leggett: {\it in Modern Trends in the Theory of Condensed Matter}, 
edited by A. Pekalski and J. Przystawa (Springer, Berlin, 1980).

\bibitem{gurarieAP}
V. Gurarie and L. Radzihovsky, Ann. Phys. (N.Y.) {\bf 322}, 2 (2007).

\bibitem{GRA}
V. Gurarie, L. Radzihovsky, and A.V. Andreev, Phys. Rev. Lett. {\bf 94}, 230403 (2005).

\bibitem{tanaka}
S. Kashiwaya and Y. Tanaka, Rep. Prog. Phys. {\bf 63}, 1641 (2000).

\bibitem{cdgm}
C. Caroli, P.G. de Gennes, and J. Matricon, Phys. Lett. {\bf 9}, 307 (1964).

\bibitem{hayashiPRL}
N. Hayashi, T. Isoshima, M. Ichioka, and K. Machida, Phys. Rev. Lett. {\bf 80}, 2921 (1998).

\bibitem{hayashiJPSJ}
N. Hayashi, M. Ichioka, and K. Machida, J. Phys. Soc. Jpn. {\bf 67}, 3368 (1998).

\bibitem{bulgac}
A. Bulgac and Y. Yu, Phys. Rev. Lett. {\bf 91}, 190404 (2003).

\bibitem{feder}
D.L. Feder, Phys. Rev. Lett. {\bf 93}, 200406 (2004).

\bibitem{mmachida1}
M. Machida and T. Koyama, Phys. Rev. Lett. {\bf 94}, 140401 (2005).

\bibitem{mmachida2}
M. Machida, Y. Ohashi, and T. Koyama, Phys. Rev. A {\bf 74}, 023621 (2006).

\bibitem{sensarma}
R. Sensarma, M. Randeria, and T.-L. Ho, Phys. Rev. Lett. {\bf 96}, 090403 (2006).

\bibitem{levin}
C.-C. Chien, Y. He, Q. Chen, and K. Levin, Phys. Rev. A {\bf 73}, 041603(R) (2006). 

\bibitem{mit1}
M.W. Zwierlein, J.R. Abo-Shaeer, A. Schirotzek, C.H. Schunck, and W. Ketterle, 
Nature {\bf 435}, 1047 (2005). 

\bibitem{mit2}
M.W. Zwierlein, A. Schirotzek, C.H. Schunck, and W. Ketterle, 
Science {\bf 311}, 492 (2006).

\bibitem{virtanen}
S.M.M. Virtanen and M.M. Salomaa, Phys. Rev. B {\bf 60}, 14581 (1999).

\bibitem{ktanaka}
K. Tanaka, I. Robel, and B. Jank\'{o}, Proc. Natl. Acad. Sci. U.S.A. {\bf 99}, 5233 (2002).

\bibitem{mizushima05}
T. Mizushima, K. Machida, and M. Ichioka, Phys. Rev. Lett. {\bf 95}, 117003 (2005).

\bibitem{takahashi}
M. Takahashi, T. Mizushima, M. Ichioka, and K. Machida, Phys. Rev. Lett. {\bf 97} 180407 (2006).

\bibitem{hu1}
H. Hu, X.-J. Liu, and P.D. Drummond, Phys. Rev. Lett. {\bf 98}, 060406 (2007).

\bibitem{hu2}
H. Hu and X.-J. Liu, Phys. Rev. A {\bf 75}, 011603(R) (2007).

\bibitem{suzuki}
K. M. Suzuki, T. Mizushima, M. Ichioka, and K. Machida, Phys. Rev. A {\bf 77} 063617 (2008).

\bibitem{ivanov}
D. A. Ivanov, Phys. Rev. Lett. {\bf 86}, 268 (2001). 

\bibitem{kopnin}
N.B. Kopnin and M.M. Salomaa, Phys. Rev. B {\bf 44}, 9667 (1991).

\bibitem{volovik}
G.E. Volovik, JETP Lett. {\bf 70}, 609 (1999).

\bibitem{yip1}
C.-K. Lu and S.-K. Yip, Phys. Rev. B {\bf 78}, 132502 (2008)


\bibitem{fujimoto}
S. Fujimoto, Phys. Rev. B {\bf 77} 220501 (2008).

\bibitem{sato}
M. Sato and S. Fujimoto, Phys. Rev. B {\bf 79}, 094504 (2009).

\bibitem{jackiw}
R. Jackiw and C. Rebbi, Phys. Rev. D {\bf 13}, 3398 (1976).

\bibitem{ssh}
R. Jackiw and J.R. Schrieffer, Nucl. Phys. {\bf B190}, 254 (1981).

\bibitem{machida1}
K. Machida and H. Nakanishi, Phys. Rev. B {\bf 30} 122 (1984).

\bibitem{machida2}
K. Machida and M. Fujita, Phys. Rev. B {\bf 30}, 5284 (1984).

\bibitem{tewari}
S. Tewari, S. Das Sarma, and D.-H. Lee, Phys. Rev. Lett. {\bf 99}, 037001 (2007). 

\bibitem{stone2}
M. Stone and R. Roy, Phys. Rev. B {\bf 69}, 184511 (2004).

\bibitem{gurarie}
V. Gurarie and L. Radzihovsky, Phys. Rev. B {\bf 75}, 212509 (2007).

\bibitem{majorana}
{\it ``Ettore Majorana''}, ed. by G.F. Bassani and the Council of
the Italian Physical Society 
(Springer, Heidelberg, 2006).

\bibitem{stone}
M. Stone and S.-B. Chung, Phys. Rev. B {\bf 73}, 014505 (2006).

\bibitem{review}
C. Nayak, S.H. Simon, A. Stern, M. Freedman, S. Das Sarma, Rev. Mod. Phys. {\bf 80}, 1083 (2008).

\bibitem{kitaev}
A. Kitaev, Ann. Phys. (N.Y.) {\bf 303}, 2 (2003).

\bibitem{freedman}
M. Freedman, M. Larsen, and Z. Wang, Commun. Math. Phys. {\bf 227}, 605 (2003).

\bibitem{bloch}
I. Bloch, J. Dalibard, and W. Zwerger, Rev. Mod. Phys. {\bf 80}, 885 (2008). 


\bibitem{mizushima}
T. Mizushima, M. Ichioka, and K. Machida, Phys. Rev. Lett. {\bf 101}, 150409 (2008).

\bibitem{chen}
Q. Chen, J. Stajic, S. Tan, and K. Levin, Phys. Rep. {\bf 412}, 1 (2005).

\bibitem{weinberg}
E.J. Weinberg, Phys. Rev. D {\bf 24}, 2669 (1981). 

\bibitem{rossi}
R. Jackiw and P. Rossi, Nucl. Phys. {\bf 190}, 681 (1981). 

\bibitem{botelho}
S. S. Botelho and C.A.R. S\'{a} de Melo, J. Low Temp. Phys. {\bf 140}, 409 (2005).

\bibitem{randeria}
M. Randeria, J.-M. Duan, and L.-Y. Shieh, 
Phys. Rev. B {\bf 41}, 327 (1990).

\bibitem{sauls}
J.A. Sauls and M Eschrig, New J. Phys. {\bf 11}, 075008 (2009).

\bibitem{stone08}
M. Stone and I. Anduagaa, Ann. Phys. (N.Y.) {\bf 323}, 2 (2008).

\bibitem{matsumoto2}
M. Matsumoto and R. Heeb, Phys. Rev. B {\bf 65}, 014504 (2001).

\bibitem{mizushimaJPS}
T. Mizushima, M. Ichioka, and K. Machida, J. Phys. Soc. Jpn. {\bf 76}, 104006 (2007).

\bibitem{sarma}
M. Cheng, R.M. Lutchyn, V. Galitski, and S. Das Sarma,  Phys. Rev. Lett. {\bf 103}, 107001 (2009).

\bibitem{kraus}
Y.E. Kraus, A. Auerbach, H.A. Fertig, and S.H. Simon, Phys. Rev. Lett. {\bf 101}, 267002 (2008). 

\bibitem{kraus2}
Y.E. Kraus, A. Auerbach, H.A. Fertig, and S.H. Simon, Phys. Rev. B. {\bf 79}, 134515 (2009). 

\bibitem{mizushima09}
T. Mizushima and K. Machida, unpublished.

\bibitem{takigawa}
M. Takigawa, M. Ichioka, K. Machida, and M. Sigrist, Phys. Rev. B {\bf 65}, 014508 (2001). 

\bibitem{massignan}
P. Massignan, A. Sanpera, and M. Lewenstein, arXiv:0908.4568.


\bibitem{baraban}
M. Baraban, G. Zikos, N. Bonesteel, and S. H. Simon, Phys. Rev. Lett. {\bf 103}, 076801 (2009).

\bibitem{simon}
Y. Tserkovnyak and S.H. Simon, Phys. Rev. Lett. {\bf 90}, 016802 (2003). 


\bibitem{lahtinen}
V.Lahtinena, G. Kellsb, A. Carolloc, T. Stittd, J. Valab, and J.K. Pachosa, 
Ann. Phys. (N.Y.) {\bf 323}, 2286 (2008).

\bibitem{bonderson}
P. Bonderson, Phys. Rev. Lett. {\bf 103}, 110403 (2009).

\bibitem{tewari07}
S. Tewari, S. Das Sarma, C. Nayak, C. Zhang, and P. Zoller, Phys. Rev. Lett. {\bf 98}, 010506 (2007).


\end{thebibliography}
\end{document}